\documentclass[twocolumn, twocolappendix,textcolor]{aastex701}
\usepackage{comment}

\begin{document}

\title{A Cyanopolyyne-rich but COM-poor Massive Protostar: \\The First Hot Carbon Chain Chemistry Source G28.28-0.36}

\author[orcid=0000-0003-4402-6475]{Kotomi Taniguchi}
\affiliation{National Astronomical Observatory of Japan, National Institutes of Natural Sciences, 2-21-1, Osawa, Mitaka, Tokyo 181-8588, Japan}
\affiliation{Astronomical Science Program, The Graduate University for Advanced Studies, SOKENDAI, 2-21-1 Osawa, Mitaka, Tokyo 181-8588, Japan}
\email[show]{kotomi.taniguchi@nao.ac.jp}

\author[orcid=0000-0003-0769-8627]{Masao Saito}
\affiliation{National Astronomical Observatory of Japan, National Institutes of Natural Sciences, 2-21-1, Osawa, Mitaka, Tokyo 181-8588, Japan}
\affiliation{Astronomical Science Program, The Graduate University for Advanced Studies, SOKENDAI, 2-21-1 Osawa, Mitaka, Tokyo 181-8588, Japan}
\email{masao.saito@nao.ac.jp} 

\author[orcid=0000-0003-1602-6849]{Prasanta Gorai}
\affiliation{Rosseland Centre for Solar Physics, University of Oslo, PO Box 1029 Blindern, 0315 Oslo, Norway}
\affiliation{Institute of Theoretical Astrophysics, University of Oslo, PO Box 1029 Blindern, 0315 Oslo, Norway}
\affiliation{Universität Heidelberg, Zentrum f\"{u}r Astronomie, Institut f\"{u}r Theoretische Astrophysik, Albert-Ueberle-Str. 2, 69120 Heidelberg, Germany}
\email{prasanta.astro@gmail.com} 

\author[orcid=0000-0002-9148-1625]{Olli Sipil\"{a}}
\affiliation{Max-Planck-Institut f\"{u}r Extraterrestrische Physik, Gie{\ss}enbachstrasse 85748 Garching bei M\"{u}nchen, Germany}
\email{osipila@mpe.mpg.de} 

\author[orcid=0000-0001-8058-8577]{Kazuhito Dobashi}
\affiliation{Tokyo Gakugei University, Koganei, Tokyo, 184-8501, Japan}
\email{dobashi@u-gakugei.ac.jp} 

\author[0000-0003-1481-7911]{Paola Caselli}
\affiliation{Max-Planck-Institut f\"{u}r Extraterrestrische Physik, Gie{\ss}enbachstrasse 85748 Garching bei M\"{u}nchen, Germany}
\email{caselli@mpe.mpg.de}

\author[orcid=0000-0001-5921-5784]{Tirupati Kumara Sridharan}
\affiliation{National Radio Astronomy Observatory, 520 Edgemont Road, Charlottesville, Virginia, USA}
\email{tksridha@nrao.edu} 

\author[orcid=0000-0002-1054-3004]{Tomomi Shimoikura}
\affiliation{Otsuma Women’s University Chiyoda, Tokyo, 102-8357, Japan}
\email{ikura@otsuma.ac.jp} 

\author[0000-0002-3389-9142]{Jonathan C. Tan}
\affiliation{Department of Physics and Astronomy, Chalmers University of Technology, 412 93  Gothenburg, Sweden}
\affiliation{Department of Astronomy, University of Virginia, Charlottesville, VA 22904, USA}
\email{jctan.astro@gmail.com}

%\collaboration{all}{The Terra Mater collaboration}

\begin{abstract}
We present molecular emission line data from the massive young stellar object (MYSO) G28.28-0.36 (G28.28) obtained with the Atacama Large Millimeter/submillimeter Array Band 3.
Cyanopolyynes (HC$_3$N and HC$_5$N) and three complex organic molecules (COMs; CH$_3$OH, CH$_3$CN, and CH$_3$CHO) are detected from the MYSO G28.28.
In addition, strong emission regions of cyanopolyynes are identified between G28.28 and a nearby ultracompact \ion{H}{2} region.
The HC$_5$N emission is coincident with the dust continuum peak, where an excitation temperature of 100 K is derived from CH$_3$CN. 
These results suggest that the Hot Carbon Chain Chemistry (HCCC) mechanism produces cyanopolyynes in the hot region around G28.28.
We find that G28.28 exhibits a unique chemical feature: cyanopolyynes are abundant, but COMs are deficient, unlike the other MYSOs studied previously.
These results imply that G28.28 is a counterpart of the Warm Carbon Chain Chemistry (WCCC) low-mass source L1527.
G28.28 is the first HCCC source identified so far. 
\end{abstract}



\section{Introduction}\label{sec:intro}

Chemical compositions around protostars are essential because they provide information on protostellar evolutionary stages, physical conditions (gas density, temperature), and the environments in which protostars formed.
Chemical complexities in protostellar envelopes have been discovered, but their origins remain unclear.  
Saturated complex organic molecules (COMs), which consist of six atoms or more at least one of which is carbon \citep{2009ARA&A..47..427H}, are abundant in the hot ($T \geq 100$ K) gas around protostellar cores. 
This chemical feature is known as ``hot cores'' and ``hot corinos'' for high-mass and low-mass protostars, respectively.

Unsaturated carbon-chain species have been known to be abundant in cold gas at the starless core stage \citep[e.g.,][]{1992ApJ...392..551S}.
These species mainly form via the bottom-up process starting from ionized carbon (C$^+$) and atomic carbon (C) in the cold gas \citep[][and references therein]{2024ApSS.369...34T}.
Another carbon-chain formation process has been found around the low-mass protostar L1527 in the Taurus star-forming region and was named {\it {Warm Carbon Chain Chemistry}} \citep[WCCC;][]{2008ApJ...672..371S}.
The detected carbon-chain emission comes from lukewarm envelopes around protostars with temperatures around 20 -- 30 K.
In the WCCC mechanism, the first step for carbon-chain formation is the gas-phase reaction between methane (CH$_4$), which sublimates from dust grains at $\sim 25$ K, and C$^+$ \citep{2008ApJ...681.1385H}.
Recently, it has been found that the WCCC mechanism occurs around intermediate-mass protostars \citep{2024AA...692A..65T}.

So far, carbon-chain chemistry has been studied mainly in nearby low-mass star-forming regions.
The chemical evolution using carbon-chain species in high-mass star-forming regions was investigated with survey observations toward high-mass starless cores (HMSCs) and high-mass protostellar objects (HMPOs) using the Nobeyama 45\,m radio telescope by \citet{2018ApJ...854..133T, 2019ApJ...872..154T}.
They found that HC$_3$N forms efficiently in lukewarm gas around early-stage HMPOs, which are still deeply embedded in dense gas and dust.

For later-stage high-mass sources, \citet{2014MNRAS.443.2252G} conducted survey observations of the HC$_5$N ($J=12-11$; $E_{\rm{up}} = 9.96$ K) line toward 79 hot cores associated with the 6.7 GHz methanol masers using the Tidbinbilla 34\,m telescope. 
The line was detected from 35 sources. 
However, their detection does not ensure that the HC$_5$N emission comes from lukewarm envelopes and/or inner hot cores, because the detected line can be excited even in cold gas.
\citet{2017ApJ...844...68T} observed four massive young stellar objects (MYSOs) containing hot cores, which were selected from the source list of \citet{2014MNRAS.443.2252G}, using the Green Bank 100\,m and Nobeyama 45\,m radio telescopes.
They detected high-$J$ lines ($E_{\rm{up}} = 63-100$ K) that could not be excited in cold gas and obtained rotational temperatures of $\sim 20-25$ K.
These results suggest that the WCCC mechanism produces HC$_5$N even around MYSOs.
These single-dish observations cannot resolve a hot core ($>100$ K) region, and it was unclear whether HC$_5$N exists in the hot core region yet.
Follow-up observations show that G28.28-0.36 (hereafter G28.28) is the most HC$_5$N-rich but COM-poor MYSO among their sample \citep{2018ApJ...866..150T, 2021ApJ...908..100T}.

Interferometric observations toward G28.28 were conducted using the Karl G. Jansky Very Large Array (VLA) in the Ka-band \citep{2018ApJ...866...32T}.
Some strong emission regions of cyanopolyynes (HC$_3$N and HC$_5$N) were found between G28.28 and a nearby ultracompact \ion{H}{2} (UC\ion{H}{2}) region, and these regions overlap with the 450 $\mu$m continuum emission. 
\citet{2018ApJ...866...32T} suggested that these regions harbor low-mass or intermediate-mass protostellar cores and the WCCC mechanism likely produces cyanopolyynes.
However, the VLA observations did not cover high-$J$ lines of cyanopolyynes, and therefore, they could not trace the spatial distribution of a potential hot component of HC$_5$N.

\citet{2023ApJS..267....4T} analyzed Atacama Large Millimeter/submillimeter Array (ALMA) Band 3 data toward five MYSOs. 
The HC$_5$N ($J=35-34$; $E_{\rm{up}} = 80.5$ K) line was detected from three MYSOs where large nitrogen (N)-bearing COMs (CH$_2$CHCN and CH$_3$CH$_2$CN) were detected. 
They confirmed that the detected HC$_5$N emission arises from hot-core regions with temperatures above 100 K, where HC$_5$N sublimates from dust grains via thermal desorption.
The formation process of cyanopolyynes in hot cores was named {\it {Hot Carbon Chain Chemistry}} \citep[HCCC;][]{2023ApJS..267....4T}.
In the HCCC mechanism, cyanopolyynes form in lukewarm gas by the neutral-neutral reactions of ``C$_{2n}$H$_2$ + CN'' and accumulate into ice mantles ($T < 100$ K).
After the temperature exceeds their sublimation temperatures ($> 100$ K), they sublimate into the gas phase and reach their peak abundances in hot-core regions.
Thus, their emission peaks are expected to be coincident with the continuum peaks associated with MYSOs.
This study demonstrates that ALMA Band 3 observations are suitable for studying carbon-chain chemistry in hot regions.

The previous studies show that the chemical diversity arises not only in low-mass protostars but also in high-mass protostars.
\citet{2024ApSS.369...34T} summarizes the chemical diversity based on stellar mass (see Figure 9 in their paper).
The MYSOs studied by \citet{2023ApJS..267....4T} correspond to hybrid-type sources in which both cyanopolyynes and a variety of COMs coexist.
However, no MYSO corresponding to the HCCC source, where cyanopolyynes are abundant but COMs are deficient in hot gas, has been discovered yet\footnote{The ``HCCC mechanism'' refers to the formation process of carbon-chain species that can proceed in both hybrid-type sources and HCCC sources. Here, we distinguish between hybrid-type and HCCC sources based on the categorization by \citet{2024ApSS.369...34T}. The categorization was proposed as counterparts of the low-mass case.}.
The MYSO G28.28 is the most promising candidate of an HCCC source \citep{2024ApSS.369...34T}, because COMs are much less abundant than in the other MYSOs \citep{2018ApJ...866..150T}.
To confirm that G28.28 is a {\it {bona fide}} HCCC source, we need high angular resolution and high sensitivity data that spatially resolve molecular emission at a central hot core region.
In particular, information on the spatial distributions of the high-$J$ HC$_5$N lines is necessary.

In this paper, we present ALMA Band 3 data toward G28.28 \citep[$d=3$ kpc;][]{2014MNRAS.443.2252G}.
The MYSO G28.28 has been identified as an extended green object (EGO)\footnote{\citet{2008AJ....136.2391C} identified extended 4.5 $\mu$m sources (Extended Green Objects; EGOs) based on the Galactic Legacy Infrared Mid-Plane Survey Extraordinaire \citep[GLIMPSE;][]{2003PASP..115..953B}.} associated with a likely MYSO outflow candidate \citep{2008AJ....136.2391C}, and with a 6.7 GHz Class II CH$_3$OH maser \citep{2009ApJ...702.1615C}.
The UC\ion{H}{2} region G28.29-0.36 is located at 18$^{\rm {h}}$44$^{\rm {m}}$15\fs09, -4\degr17\arcmin54\farcs9 (J2000), corresponding to the east side of the MYSO G28.28 \citep{2009ApJ...702.1615C}. 

This paper is organized as follows.
Section \ref{sec:obs} explains details of the observations with ALMA and the data reduction method.
We present the results of a continuum image and moment 0 maps of detected molecular lines as well as their analyses in Section \ref{sec:res}.
The properties of the strong cyanopolyyne emission regions are discussed in Section \ref{sec:dis1}.
We compare the chemical compositions in some representative positions (Section \ref{sec:dis2}) and between G28.28 and the other MYSOs studied by \citet{2023ApJS..267....4T} (Section \ref{sec:dis3}).
In Section \ref{sec:dis5}, we summarize the chemical features of G28.28.
Our main conclusions are summarized in Section \ref{sec:con}.

\section{Observations \& Data Reduction}\label{sec:obs}

Observations were conducted during the ALMA Cycle 10 (ID: 2023.1.00467.S., PI: Kotomi Taniguchi).
The observations with the 12-m array ran in 2023 December and 2024 January, and those with the 7-m array were conducted in 2024 March.
The center of the field of view (FoV) is set at the MYSO G28.28, whose coordinates are 18$^{\rm {h}}$44$^{\rm {m}}$13\fs3, -4\degr18\arcmin03\farcs3 (J2000).
The synthesized beam size of the 12-m array is $1.54'' \times 1.25''$ (mean value is $1.39''$), corresponding to $\sim 0.02$ pc at the source distance (3 kpc).
The maximum recoverable scale and FoV size of the 12-m array are $19.2''$ and $56.5''$, respectively.
The synthesized beam size of the 7-m array is $12.3''$.
The maximum recoverable scale of the 7-m array and FoV size are $83.3''$ and $96.8''$, respectively.

Table \ref{tab:spw} summarizes the spectral window setup.
Spectral windows targeting the molecular lines had a bandwidth of 62 MHz with a frequency resolution of 60.5 kHz, corresponding to a velocity resolution of $\sim0.17$ km\,s$^{-1}$ in the 3\,mm band.
One spectral window with a 2-GHz bandwidth was set at 108.492 GHz to improve the calibration.

We carried out data reduction using the Common Astronomy Software Applications package \citep[CASA;][]{2022PASP..134k4501C}.
We used the pipeline version 2023.1.0.124 with CASA 6.5.4.9. 
The continuum image and data cubes were created using the CASA tclean task, combining data of the 12-m and 7-m arrays.
The pixel size was set at $0.3''$.
The Briggs weighting (robust = 0.5) was applied.
The final resulting beam size is $1.4'' \times 1.1''$, corresponding to 0.020 pc $\times$ 0.016 pc at the source distance.

\begin{deluxetable}{lcc}
\tablewidth{0pt}
\tablecaption{Spectral window setup \label{tab:spw}}
\tablehead{\colhead{Main Target Line} & \colhead{Center Frequency} & \colhead{Bandwidth} \\
\colhead{} & \colhead{(GHz)} & \colhead{(MHz)}
}
\startdata 
HC$_5$N ($36-35$) & 95.840 & 62 \\
CH$_3$CHO ($5_{0,5}-4_{0,4}$) & 95.953 & 62 \\
CH$_3$OH ($2_{1,2}-1_{1,1}$) & 95.904 & 62 \\
CH$_3$OH ($2_{1,1}-1_{1,0}$) & 96.745 & 62 \\
CH$_3$OCHO ($8_{3,6}-7_{3,5}$) & 98.596 & 62 \\
HC$_5$N ($37-36$) & 98.502 & 62 \\
CH$_3$CH$_2$CN ($12_{2,10}-11_{2,9}$) & 108.929 & 62 \\
HC$_3$N ($12-11$) & 109.162 & 62 \\
CH$_3$CN ($J=6-5$) & 110.363 & 62 \\
Continuum & 108.492 & 2000 \\
\enddata
%\tablenotetext{a}{Values are deconvolved from the beam.}
\end{deluxetable}

\section{Results \& Analyses}\label{sec:res}

\subsection{Continuum emission}\label{ssec:cont}

\begin{figure*}[th!]
 \centering
  \includegraphics[width=0.8\textwidth, bb = 0 5 300 140]{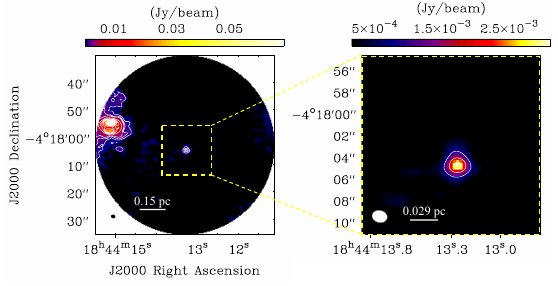}
\caption{Continuum ($\lambda = 2.75$ mm) images toward the MYSO G28.28. The left panel shows the entire image, and the right panel shows a close-up image centered on the MYSO G28.28. The strong emission region at the east side in the left panel corresponds to the UC\ion{H}{2} region G28.29-0.36. The noise level ($1\,\sigma$) is 0.18 mJy\,beam$^{-1}$. The contour levels in the left panel are 5, 10, 15, 35, 55, 75, 100, 150, 200, 250, 300, 350 $\sigma$, and those in the right panel are 5, 10, 15 $\sigma$.
\label{fig:continuum}}
\end{figure*}

Figure \ref{fig:continuum} shows the continuum ($\lambda = 2.75$ mm, corresponding to 108.9 GHz) image combining the 12-m and 7-m arrays.
The MYSO G28.28 has been detected with a signal-to-noise (S/N) ratio well above 5\,$\sigma$.

We conducted the 2-dimensional (2D) Gaussian fitting for the continuum source of the MYSO G28.28.
The fitting results are summarized in Table \ref{tab:continuum}.
We constrained its stellar mass to $8-16$ $M_\odot$ with spectral energy distribution (SED) fitting (Figures \ref{fig:sed} and \ref{fig:sed-para}, and Table \ref{tab:sedfit} in Appendix \ref{sec:sed}).
The bolometric luminosity, core mass, and core radius are derived to be $(1.9-5.2) \times 10^4$ $L_\odot$, $100-480$ $M_\odot$, and $0.04-0.5$ pc, respectively (Table \ref{tab:sedfit} in Appendix \ref{sec:sed}).

\begin{deluxetable}{lcc}
\tablewidth{0pt}
\tablecaption{Summary of the 2D Gaussian fitting for G28.28 in the continuum image \label{tab:continuum}}
\tablehead{\colhead{Parameter} & \colhead{Value} & \colhead{Error}
}
\startdata 
      R.A. (J2000) & 18$^{\rm {h}}$44$^{\rm {m}}$13\fs262 & 0.003$^{\rm{s}}$ \\
      Dec. (J2000) & -4\degr18\arcmin04\farcs85 & $0.05\arcsec$ \\
      Major axis FWHM [$\arcsec$]\tablenotemark{a} & 1.68 & 0.16 \\
      Minor axis FWHM [$\arcsec$]\tablenotemark{a} & 1.42 & 0.18 \\
      Position angle [deg]\tablenotemark{a} & 163 & 149 \\
      Flux [mJy] & 6.85 & 0.51 \\
      Peak [mJy/beam] & 2.68 & 0.15 \\
\enddata
\tablenotetext{a}{Values are deconvolved from the beam.}
\end{deluxetable}

\subsection{Molecular lines}\label{ssec:line}

\subsubsection{Moment 0 maps}

\begin{figure*}[th!]
  \centering
  \includegraphics[bb = 10 5 460 490, scale = 1.15]{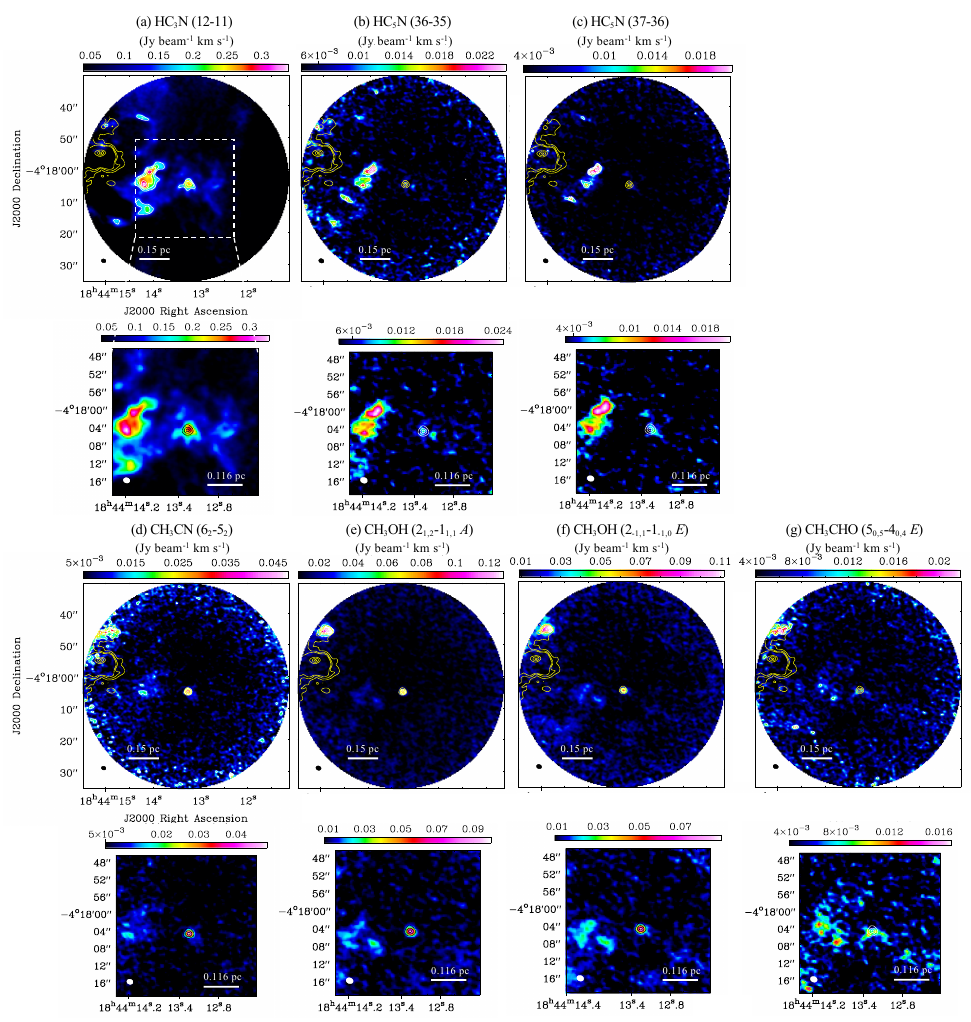} 
\caption{Moment 0 maps of the molecular lines. The upper three panels show maps of cyanopolyynes: (a) HC$_3$N ($J=12-11$), (b) HC$_5$N ($J=36-35$), and (c) HC$_5$N ($J=37-36$). The lower four panels show maps of complex organic molecules: (d) CH$_3$CN ($J_K = 6_2-5_2$), (e) CH$_3$OH ($2_{1,2}-1_{1,1}$ $A$), (f) CH$_3$OH ($2_{-1,1}-1_{-1,0}$ $E$), and (g) CH$_3$CHO ($5_{0,5}-4_{0,4}$ $E$). Color scale and white contours show moment 0 maps of each line, and yellow contours are the continuum image (5, 10, 15, 150, 250, 350\,$\sigma$). The rms noise (denoted as $\sigma$) and contour levels of each panel are as follows: (a) rms = 0.04 Jy\,beam$^{-1}$ km\,s$^{-1}$ (4, 5, 6, 7, 8\,$\sigma$); (b) rms = 0.0043 Jy\,beam$^{-1}$ km\,s$^{-1}$ (3, 4, 5\,$\sigma$); (c) rms = 0.0032 Jy\,beam$^{-1}$ km\,s$^{-1}$ (3, 4, 5, 6\,$\sigma$); (d) rms = 0.005 Jy\,beam$^{-1}$ km\,s$^{-1}$ (3, 4, 5, 6\,$\sigma$); (e) rms = 0.0077 Jy\,beam$^{-1}$ km\,s$^{-1}$ (5, 7, 9, 11, 13, 15\,$\sigma$); (f) rms = 0.009 Jy\,beam$^{-1}$ km\,s$^{-1}$ (5, 7, 9, 11\,$\sigma$); (g) rms = 0.0039 Jy\,beam$^{-1}$ km\,s$^{-1}$ (3, 4, 5\,$\sigma$). Close-up images of each panel are shown below. These regions are indicated by a white dashed square in panel (a). Color scales are adjusted for the close-up ones. The black and white contours are the continuum map same as in the right panel of Figure \ref{fig:continuum}.  
\label{fig:mom0}}
\end{figure*}

Figure \ref{fig:mom0} shows moment 0 maps of the detected molecular lines.
All of the lines have been detected at the continuum peak of the MYSO G28.28.
Panels (a) -- (c) show spatial distributions of cyanopolyynes (HC$_3$N and HC$_5$N).
The HC$_5$N lines show a weak emission peak at the G28.28 continuum peak.
The observed HC$_5$N lines have high $E_{\rm{up}}$ values ($\approx85$ K), and their detection suggests that the emission arises from warm envelopes or hot ($T>100$ K) gas.
In addition to the G28.28 continuum peak, HC$_3$N and HC$_5$N show strong emission between the MYSO G28.28 and the UC\ion{H}{2} region, even though no strong continuum peaks have been detected in the 2.75 mm continuum map (Figure \ref{fig:continuum}).
We will discuss properties of the emission regions in Section \ref{sec:dis1}.

The HC$_3$N emission also traces extended components. 
Diffuse components extending in the west-east direction centered on the MYSO G28.28 are detected.
This feature may be related to extended envelope/cloud material and/or outflow activity. 
The HC$_3$N ($J=11-10$) line has been reported as a potential tracer of low-velocity molecular outflows associated with massive protostars \citep{2025ApJ...987..197H}.
Although G28.28 has been identified as an EGO likely associated with outflow(s), the HC$_3$N moment 1 map and position-velocity diagram do not allow us to identify outflows unambiguously because of the complex velocity structure.
Additional observations of dedicated jet/outflow tracers are needed to clarify the origin of the extended HC$_3$N emission.

In the case of COMs (CH$_3$CN, CH$_3$OH, and CH$_3$CHO), the emission between the MYSO G28.28 and the UC\ion{H}{2} region is not as strong as for the cyanopolyynes, as shown in panels (d) -- (g) of Figure \ref{fig:mom0}.
Strong emission peaks are detected at the northern part of the UC\ion{H}{2} region.
These peaks marginally correspond to the 2.75 mm continuum emission peaks.
No source has yet been identified at the COM emission peaks located north of the UC\ion{H}{2} region.

Two spectral windows were set for CH$_3$CH$_2$CN ($12_{2,10}-11_{2,9}$; $E_{\rm {up}}=38.2$ K) and CH$_3$OCHO ($8_{3,6}-7_{3,5}$; $E_{\rm {up}}=27.3$ K).
However, these lines of the large COMs were not detected with the current sensitivities.
These two lines have low $E_{\rm {up}}$ values, and they should be easily detected if G28.28 is a COM-rich hot core. 
In fact, the similar $E_{\rm {up}}$ line of CH$_3$CH$_2$CN has been detected from the MYSOs studied by \citet{2023ApJS..267....4T}.
Their non-detection therefore suggests that G28.28 is deficient in large COMs; this point is quantified in Section \ref{sec:dis3}.

\subsubsection{Spectral Analyses with CASSIS}

\begin{figure}[ht!]
 \centering
  \includegraphics[bb = 0 8 200 160, width = 8cm]{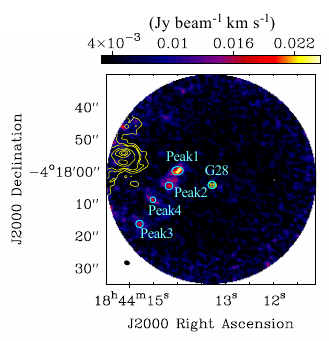} 
\caption{Positions selected for line analysis. Color scale indicates the moment 0 map of the HC$_5$N ($J=36-35$) line, and the yellow contours show the continuum emission of the 5, 10, 15, 50, 100, 200, 300 $\sigma$. Cyan circles/ellipses indicate the region for the spectral analysis. 
\label{fig:posi}}
\end{figure}

We selected positions to analyze molecular lines based on the moment 0 map of the HC$_5$N ($J=36-35$) line.
For G28.28, we double-checked the moment 0 map of CH$_3$CN because it is a typical hot-core tracer. The aperture size of $2.3\arcsec$ for G28.28 was chosen to cover molecular emission coming from the hot-core region. This aperture size corresponds to 0.03 pc at the source distance (3 kpc), and is hence reasonable for deriving the hot-core averaged column densities.
In addition to the MYSO G28.28, we identified four peaks between the MYSO G28.28 and the UC\ion{H}{2} region, as shown in cyan circles/ellipses in Figure \ref{fig:posi}.
We created the spectra by averaging the emission within these circles/ellipses.
We named these peaks Peak\,1 -- Peak\,4, respectively.
Table \ref{tab:HC5Npeak} summarizes the information on these positions.
There are no clear continuum peaks at the positions of Peak\,1 -- Peak\,4 in Figure \ref{fig:continuum}, but weak emission was detected.
The continuum fluxes at Peak\,1 -- Peak\,4 are lower than that at the MYSO G28.28 by factors of 7 -- 23. 
We calculated the H$_2$ column density and number density at each position using the 2.75 mm continuum data.
The method to derive the column densities of H$_2$, $N$(H$_2$), is explained in Appendix \ref{sec:append1}.
The values of $N$(H$_2$) and $n$(H$_2$) listed in Table \ref{tab:HC5Npeak} are the aperture-averaged values. 
When we applied the results of the 2D Gaussian fit for G28.28 (Table \ref{tab:continuum}), $N$(H$_2$), $\tau$, and $n$(H$_2$) are derived to be $3.3 \times 10^{24}$ cm$^{-2}$, 0.058, and $5.3 \times 10^7$ cm$^{-3}$, respectively, that is, the obtained values change by only a factor of a few.

\begin{deluxetable}{lccccc}
\tablewidth{0pt} 
\tablecaption{Properties of positions for molecular line analyses \label{tab:HC5Npeak}}
\tablehead{\colhead{Position} & \colhead{Coordinate (R.A., Dec.)\tablenotemark{a}} & \colhead{Size [$\arcsec \times \arcsec$]\tablenotemark{b}} & \colhead{Flux [mJy]\tablenotemark{c}} & \colhead{$N$(H$_2$) [cm$^{-2}$]\tablenotemark{d,e}} & \colhead{$n$(H$_2$) [cm$^{-3}$]\tablenotemark{e}} 
}
\startdata
      G28.28   & 18$^{\rm {h}}$44$^{\rm {m}}$13\fs278, -4\degr18\arcmin04\farcs749 & $2.3 \times 2.3$ & $77 \pm 17$ & $(1.2 \pm 0.3)\times 10^{24}$ & $(1.2 \pm 0.2) \times 10^7$ \\
      Peak\,1 & 18$^{\rm {h}}$44$^{\rm {m}}$13\fs991, -4\degr18\arcmin00\farcs392 & $3.5 \times 2.6$ & $6 \pm 2$ & $(1.4 \pm 0.4) \times 10^{23}$ & $(1.1 \pm 0.4) \times 10^6$\\
      Peak\,2 & 18$^{\rm {h}}$44$^{\rm {m}}$14\fs169, -4\degr18\arcmin05\farcs058 & $2.3 \times 2.3$ & $11 \pm 2$ & $(4.5 \pm 0.9) \times 10^{23}$ & $(4.4 \pm 0.9) \times 10^6$\\
      Peak\,3 & 18$^{\rm {h}}$44$^{\rm {m}}$14\fs778, -4\degr18\arcmin16\farcs900 & $2.3 \times 2.3$ & $5 \pm 1$ & $(2.2 \pm 0.5) \times 10^{23}$ & $(2.1 \pm 0.5) \times 10^6$\\
      Peak\,4 & 18$^{\rm {h}}$44$^{\rm {m}}$14\fs501, -4\degr18\arcmin09\farcs433 & $1.7 \times 1.7$ & $3 \pm 2$ & $(3.2 \pm 1.9) \times 10^{23}$ & $(4.1 \pm 2.5) \times 10^6$ \\
\enddata
\tablenotetext{a}{J2000 coordinates.}
\tablenotetext{b}{The major and minor axes of the chosen circle/elliptical aperture displayed in Figure \ref{fig:posi}.}
\tablenotetext{c}{Flux values were measured in the continuum image.}
\tablenotetext{d}{Assuming the optically thin dense gas case \citep[$\beta = 1.5$;][]{2024AA...687A.163B}. Details are described in Appendix \ref{sec:append1}. }
\tablenotetext{e}{The listed values are aperture-averaged values.}
\end{deluxetable}

We analyzed molecular line data with the Markov Chain Monte Carlo (MCMC) method using the CASSIS software \citep{2015sf2a.conf..313V}.
We assumed the local thermodynamic equilibrium (LTE) condition.
This assumption is applicable because the gas densities at the five positions (Table \ref{tab:HC5Npeak}) are comparable to the critical densities of the molecular lines ($\approx 10^5-10^6$ cm$^{-3}$).
The line parameters used for the fitting are summarized in Table \ref{tab:line} in Appendix \ref{sec:append2}.
In the analyses, we fixed the emission size to the extraction region adopted for each position, as listed in Table \ref{tab:HC5Npeak}. 
We conducted 2D Gaussian fitting for the G28.28 peak in all of the moment 0 maps, except for the HC$_5$N ($J=36-35$) line, and confirmed that the emission regions of molecular lines are comparable to or larger than the chosen aperture size ($2.3\arcsec$). 
Regarding Peak\,1 -- Peak\,4 and the HC$_5$N ($J=36-35$) line at G28.28, we confirmed that all of the molecular emission regions fill the beam sizes in their moment 0 maps.  
In the CASSIS software, the beam filling factor of $\theta_{\rm {source}}^2/(\theta_{\rm {source}}^2 + \theta_{\rm {beam}}^2)$ is applied. Here, $\theta_{\rm {source}}$ and $\theta_{\rm {beam}}$ are the source size and the synthesized beam, respectively.
We fixed the source sizes at the aperture sizes for creating spectra listed in Table \ref{tab:HC5Npeak}.
The derived column densities should be regarded as aperture-averaged values.

We analyzed CH$_3$CN to obtain column densities and excitation temperatures from its $K$-ladder lines ($K=0-3$).
The excitation temperature derived from CH$_3$CN is a good thermometer for hot dense molecular cores \citep{2014ApJ...786...38H, 2014ApJ...788..187H, 2023ApJ...950...57T}.
Then, its excitation temperatures were applied to the other molecular analyses, except for CH$_3$OH.
We derived column densities and excitation temperatures of CH$_3$OH by fitting its four lines.
In the analysis of the CH$_3$OH lines, we excluded lines with non-Gaussian profiles from the fitting. 

\begin{figure*}[ht!]
 \centering
  \includegraphics[bb = 0 5 530 260, scale = 0.9]{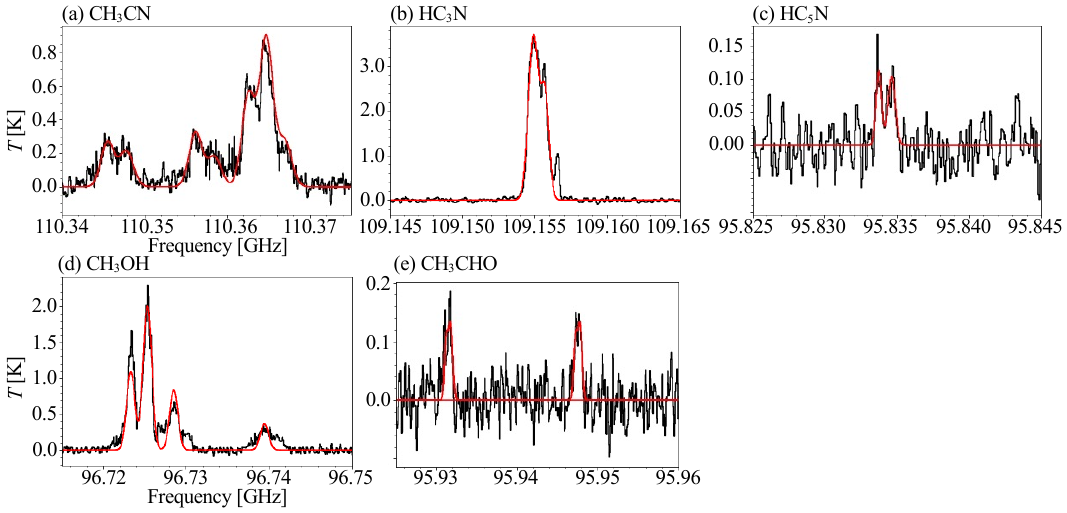} 
\caption{Spectra at the G28.28 continuum peak; (a) CH$_3$CN, (b) HC$_3$N, (c) HC$_5$N, (d) CH$_3$OH, and (e) CH$_3$CHO.  Black and red lines indicate the observed spectra and the best-fitting models by CASSIS, respectively.
\label{fig:specG28}}
\end{figure*}

Figure \ref{fig:specG28} shows the spectra at the MYSO G28.28 as examples.
The spectra at Peak\,1 -- Peak\,4 are presented in Figures \ref{fig:specP12} and \ref{fig:specP34} in Appendix \ref{sec:append2}.
The black and red lines indicate the observed spectra and the best-fitting model, respectively.
We applied a two-velocity component fit, except for CH$_3$OH, at the MYSO G28.28.
Table \ref{tab:mcmc} summarizes the parameters obtained from the MCMC analysis.
Since the HC$_5$N ($J=37-36$) line shows an S/N ratio lower than 4, we excluded it from the fitting.
The optical depths of these molecular lines are small ($\tau \approx 0.003-0.05$).

We conducted additional tests by changing emission (source) sizes to investigate how much the derived column densities are affected by uncertainties in emission sizes ($i.e.,$ beam filling factors). We changed the emission size from the aperture sizes (Table \ref{tab:HC5Npeak}) to $1.3\arcsec$ (the synthesized beam size). Even if the emission size is reduced, all of the molecular lines remain optically thin ($\tau \ll 0.1$), and the column densities increase by less than a factor of two.
Hence, the aperture-averaged column densities are not significantly affected by uncertainties in the beam filling factor.

The excitation temperature derived toward the MYSO G28.28 is 100 K, whereas those derived at Peak\,1 -- Peak\,4 show much lower values (25 -- 30 K).
The line widths at the MYSO G28.28 are wider than those at the other positions. 
These results indicate that the MYSO G28.28 harbors hotter gas compared to the other positions. 

\begin{deluxetable}{lcccc}
\tablewidth{0pt} 
%\tablenum{1}
\tablecaption{Results of the spectral analyses \label{tab:positions}}
\label{tab:mcmc}
\tablehead{
\colhead{Species} & \colhead{$N$ (cm$^{-2}$)} & \colhead{$T_{\rm {ex}}$ (K)} & \colhead{FWHM (km\,s$^{-1}$)\tablenotemark{a}} & \colhead{$V_{\rm {LSR}}$ (km\,s$^{-1}$)} 
}
\startdata 
      \multicolumn{4}{l}{G\,28 (1st Component)} \\
        HC$_3$N	&	$(6.21\pm 0.05) \times 10^{13}$	& 50.2 (fixed) &	$2.23 \pm 0.17$	& $51.34	\pm	0.17$ \\
        HC$_5$N	&	$(3.4\pm 0.5)\times 10^{12}$ &	50.2 (fixed) & $1.44 \pm	0.22$	&	$51.84	\pm	0.17$ \\
        CH$_3$CN &	$(4.8 \pm 0.1)\times 10^{13}$ & $50.2 \pm 0.3$	& $5.32	\pm	0.27$ &	$50.97 \pm 0.18$ \\
        CH$_3$OH &	$(1.17\pm0.2) \times 10^{15}$ &	$13.2 \pm 0.3$	& $3.59	\pm	0.17$ &	$49.60 \pm 0.17$ \\
        CH$_3$CHO &	$(4.0 \pm 0.4) \times 10^{13}$ & 50.2 (fixed)	& $1.50	\pm	0.18$ &	$50.65 \pm 0.18$ \\
      \multicolumn{4}{l}{G\,28 (2nd Component)}\\
        HC$_3$N	& $(4.0\pm0.1) \times 10^{13}$ & 100.5 (fixed) & $1.69 \pm 0.18$ & $49.16	\pm	0.17$ \\	
        HC$_5$N	& $(3.1 \pm 0.4) \times 10^{12}$ & 100.5 (fixed) & $1.62 \pm 0.24$ &	$48.97 \pm 0.17$ \\	
        CH$_3$CN &	$(4.5 \pm 0.2) \times 10^{13}$ &	$100.5 \pm 0.3$	& $5.13	\pm	0.27$ &	$44.8 \pm 0.2$ \\
        CH$_3$OH &	... & ... & ... & ... \\
        CH$_3$CHO & $(1.7 \pm 0.1) \times 10^{14}$ & 100.5 (fixed) &	$1.89 \pm 0.18$	& $49.0 \pm	0.2$ \\
      \multicolumn{4}{l}{Peak 1}\\
       HC$_3$N	& $(8.82 \pm 0.03) \times 10^{13}$	& 30.1 (fixed) & $1.07 \pm 0.17$ 	& $49.42 \pm 0.17$ \\
       HC$_5$N	& $(3.92 \pm 0.09) \times 10^{13}$	& 30.1 (fixed) & $1.00 \pm 0.17$	& $49.50 \pm 0.17$ \\
       CH$_3$CN &	$(9.3 \pm 0.2) \times 10^{12}$	& $30.1	\pm	0.1$ & $1.08 \pm 0.17$ & $49.43	\pm	0.17$ \\
       CH$_3$OH &	$(2.95 \pm 0.06) \times 10^{14}$ &	$9.6 \pm	0.3$ & $1.80 \pm 0.17$ & $49.04	\pm	0.17$ \\
       CH$_3$CHO &	$(2.5 \pm 0.1) \times 10^{13}$ &	30.1 (fixed) & $1.39 \pm 0.17$ & $49.51 \pm 0.17$ \\
      \multicolumn{4}{l}{Peak 2}\\
      HC$_3$N	& $(1.12 \pm 0.01) \times 10^{14}$ & 30 (fixed) & $1.49 \pm 0.17$	& $49.32 \pm 0.17$ \\
      HC$_5$N	& $(2.6 \pm 0.1) \times 10^{13}$ & 30 (fixed) & $1.00	\pm	0.17$ &	$49.27 \pm 0.17$ \\
      CH$_3$CN & $(1.40 \pm 0.02) \times 10^{13}$ & $30.04 \pm 0.03$ & $1.42 \pm 0.17$ & $49.24 \pm 0.17$ \\
      CH$_3$OH &	$(1.27 \pm 0.01) \times 10^{15}$ &	$10.6 \pm 0.1$ & $1.41 \pm 0.17$ & $49.02 \pm 0.17$ \\
      CH$_3$CHO &	$(3.2 \pm 0.3) \times 10^{13}$ & 30 (fixed) & $2.2 \pm 0.4$	& $49.45 \pm 0.18$ \\
      \multicolumn{4}{l}{Peak 3}\\
      HC$_3$N	& $(6.39\pm0.02) \times 10^{13}$ & 30.1 (fixed) & $1.06 \pm 0.17$	& $46.23 \pm 0.17$ \\
      HC$_5$N	& $(2.47 \pm 0.09) \times 10^{13}$ & 30.1 (fixed)	& $1.50	\pm	0.17$ &	$46.28 \pm 0.17$ \\
      CH$_3$CN & $(7.0 \pm 0.3) \times 10^{12}$ &	$30.13 \pm 0.04$ & $1.44 \pm 0.17$ & $46.39	\pm	0.17$ \\
      CH$_3$OH & $(8.3 \pm 0.7) \times 10^{14}$ & $16.1 \pm 0.7$ &	$3.4 \pm 0.3$ &	$46.88 \pm 0.17$ \\
      CH$_3$CHO &	$(3.0 \pm 0.3) \times 10^{13}$ & 30.1 (fixed) & $1.54	\pm	0.18$ &	$46.69 \pm 0.18$ \\
      \multicolumn{4}{l}{Peak 4}\\
      HC$_3$N	& $(3.99 \pm 0.02) \times 10^{13}$ & 24.9 (fixed)	& $0.61 \pm 0.17$ & $50.24 \pm 0.17$ \\
      HC$_5$N	& $(4.10 \pm 0.09) \times 10^{13}$ & 24.9 (fixed)	& $0.59	\pm	0.17$ &	$50.36 \pm 0.17$ \\
      CH$_3$CN & $(2.5 \pm 0.5) \times 10^{12}$ &	$24.9 \pm 0.1$	& $0.42	\pm	0.17$ &	$50.34 \pm 0.17$ \\
      CH$_3$OH & $(2.17 \pm 0.04) \times 10^{14}$ & $10.2	\pm	0.1$ & $2.30 \pm 0.17$ & $49.10	\pm	0.17$ \\
      CH$_3$CHO & ... & ... & ... & ... \\  
\enddata
\tablenotetext{a}{Errors include the system uncertainties originating from the velocity resolution of the spectral window setup (0.168 km\,s$^{-1}$).} 
%\tablenotetext{b}{The first and second components of the CS line are $\Delta\,V_{\rm {LSR}} \approx -0.68$ km\,s$^{-1}$ and 0.1 km\,s$^{-1}$, respectively.} 
\end{deluxetable}

\section{Discussion}\label{sec:dis}

\subsection{Properties of the cyanopolyyne emission regions between the MYSO G28.28 and the UC\ion{H}{2} region} \label{sec:dis1}

We found strong emission regions of cyanopolyynes (HC$_3$N and HC$_5$N) between the MYSO G28.28 and the UC\ion{H}{2} region, $i.e.,$ Peak\,1 -- Peak\,4 (Figure \ref{fig:posi}). 
We discuss properties of these regions and the origin of the cyanopolyyne emission in this subsection.

These emission regions were also detected in the VLA Ka-band observations \citep{2018ApJ...866...32T}.
In the VLA observations, only the low-$E_{\rm{up}}$ lines were covered; 4.4 K and 13.4 K for HC$_3$N and HC$_5$N, respectively.
These lines have low critical densities ($\approx 10^4$ cm$^{-3}$).
They spatially correlate with the 450 $\mu$m continuum emission. 
\citet{2018ApJ...866...32T} proposed that these regions harbor low-mass or intermediate-mass protostellar cores and that the WCCC mechanism works. 

We found that the 2.75 mm continuum fluxes at the HC$_5$N peak positions are weaker than that at the MYSO G28.28 (Table \ref{tab:HC5Npeak}), which supports that these positions do not contain high-mass protostellar cores.
The derived $n$(H$_2$) values at Peak\,1 -- Peak\,4 are higher than $10^6$ cm$^{-3}$ (Table \ref{tab:HC5Npeak}), indicating dense gas.
As we point out in Appendix \ref{sec:append1}, our assumption that the length of the core is equal to the beam size could overestimate $n$(H$_2$) at these four positions. However, the detection of the lines with high critical densities ($10^5-10^6$ cm$^{-3}$) supports that the gas densities at Peak\,1 -- Peak\,4 are at least moderate values ($\sim10^5$ cm$^{-3}$).
Together with the detection of high-$E_{\rm {up}}$ HC$_5$N lines, this suggests that these peaks trace moderately dense and warm regions. 
They may be associated with low- or intermediate-mass protostellar cores, although the current data do not prove the presence of embedded protostars.

The derived excitation temperatures of CH$_3$CN at Peak\,1 -- Peak\,4 are $\sim 25 - 30$ K (Table \ref{tab:mcmc}), corresponding to the temperature at which the WCCC mechanism works efficiently.
Thus, the observed temperatures are consistent with conditions under which WCCC-like chemistry can enhance HC$_5$N.

We consider the possibility of external heating from the nearby UC\ion{H}{2} region, which could contribute to exciting the high-$J$ lines.
The external UV radiation could be shielded at the central cores of Peak\,1 -- Peak\,4, where the H$_2$ number densities are on the order of $10^6$ cm$^{-3}$ \citep{2011ApJ...739...63J}. 
However, the penetration of the UV radiation strongly depends on the density structure of molecular clouds. 
Thus, with the current datasets, we cannot completely exclude a contribution from external heating by the nearby UC\ion{H}{2} region.

\subsection{Comparison of chemical compositions among the five representative positions} \label{sec:dis2}

\begin{figure*}[ht!]
 \centering
  \includegraphics[bb = 10 10 430 200, scale = 0.8]{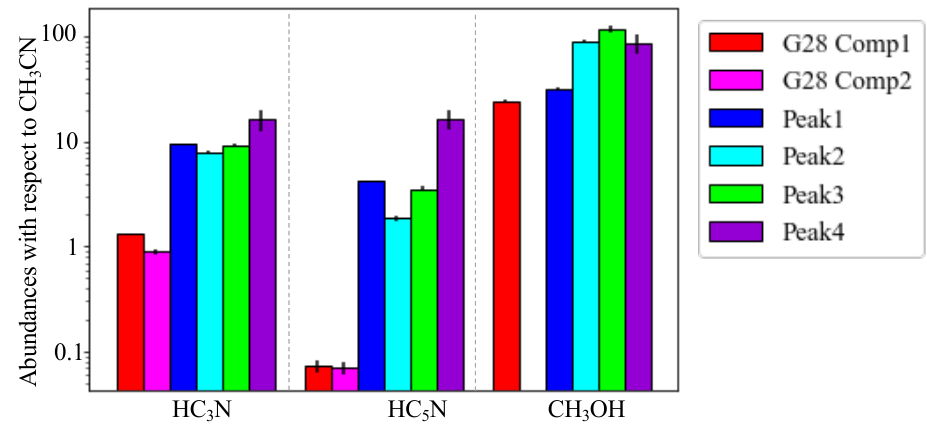} 
\caption{Comparisons of column density ratios with respect to CH$_3$CN at the five positions. The error bars indicate the standard deviation.
\label{fig:abundance}}
\end{figure*}

We compare the chemical compositions among the five positions (G28.28 and Peak\,1 -- Peak\,4) in this subsection.
We calculate and compare the column density ratios of [$X$]/[CH$_3$CN], where $X$ represents molecular species except for CH$_3$CN.
Since CH$_3$CN has been detected at all of the positions and its column densities are the most reliable due to its four-line fitting, we chose this species as the 
reference.
The beam filling factor does not affect the molecular emission ratios if the emission comes from similar regions and their lines are optically thin. In Section \ref{ssec:line}, we confirmed that all of the molecular lines are optically thin even if the emission size is smaller than the applied aperture sizes. Thus, the aperture-averaged column densities can be used for this comparison. We exclude CH$_3$CHO because of large uncertainties in emission sizes. Moreover, unlike the cyanopolyynes, the detected CH$_3$CHO lines have a low upper-state energy ($E_{\rm {up}}=13.9$ K), and their emission may contain contributions from the outer envelopes. Hence, the emission regions of the CH$_3$CHO lines could differ from those of the CH$_3$CN lines that predominantly trace the hot core region.
Figure \ref{fig:abundance} shows the column density ratios with respect to CH$_3$CN among the five positions.
Note that the fitting result of the CH$_3$OH spectra at Peak\,3 is tentative and its accuracy is low. We exclude it from the detailed discussion.

The column density ratios of HC$_3$N/CH$_3$CN and HC$_5$N/CH$_3$CN at G28.28 are $\sim 0.9-1.3$ and $\sim0.07$, respectively.
We compared the column density ratio of HC$_5$N/CH$_3$CN at G28.28 to the chemical network simulations presented by \citet{2023ApJS..267....4T}. We found that the observed ratio agrees with the model when the temperature reaches the sublimation temperatures of HC$_5$N and CH$_3$CN.
Hence, HC$_5$N is formed via the Hot Carbon Chain Chemistry (HCCC) mechanism at G28.28.
This is consistent with the derived excitation temperatures of CH$_3$CN (100.5 K; Table \ref{tab:mcmc}).

The ratios of HC$_3$N/CH$_3$CN and HC$_5$N/CH$_3$CN at the four HC$_5$N peak positions (Peak\,1 -- Peak\,4) are $\sim8-16$ and $2-16$, respectively.
Thus, cyanopolyynes at the MYSO G28.28 show lower ratios compared to the HC$_5$N peak positions by more than one order of magnitude, which means that their chemistry at G28.28 differs from that at Peak\,1 -- Peak\,4.
We found smaller differences in the HC$_3$N/CH$_3$CN ratio among Peak\,1 -- Peak\,4 compared to HC$_5$N.
This may be because the HC$_3$N emission is contaminated by outer envelopes, given that the observed HC$_3$N line has a moderate $E_{\rm{up}}$ value (34.1 K) compared to HC$_5$N (85.1 K).
The chemical composition of the natal molecular cloud could affect the column density ratio of HC$_3$N/CH$_3$CN.

Methanol (CH$_3$OH) shows lower $T_{\rm {ex}}$ values compared to CH$_3$CN (Table \ref{tab:mcmc}).
This implies that the CH$_3$OH emission comes not only from hot-core regions but also from outer envelopes.
The CH$_3$OH lines show a non-Gaussian profile at G28.28 (Figure \ref{fig:specG28}), suggesting contamination from molecular outflows and/or outer envelopes, or molecular clouds. 
Thus, the detected lines of CH$_3$OH may be contaminated by outer envelopes or natal clouds, rather than central hot gas.
This contamination may explain why the CH$_3$OH/CH$_3$CN ratios do not differ strongly among the five positions.

\subsection{Comparisons of chemical compositions between the MYSO G28.28 and other MYSOs} \label{sec:dis3}

\citet{2023ApJS..267....4T} analyzed ALMA Band\,3 data toward five MYSOs, and derived the abundances with respect to CH$_3$CN (see Figure 14 in their paper).
They reported that the HC$_5$N ($J=35-34$, $E_{\rm {up}}=80.5$ K) line is detected from four cores, and it is always associated with large COMs, which means that HC$_5$N has been detected from chemically-rich hot cores.
In this subsection, we compare the chemical compositions of the MYSO G28.28 to those presented by \citet{2023ApJS..267....4T}.

The derived HC$_5$N/CH$_3$CN ratios at G28.28 are $\approx0.07$. 
The HC$_5$N/CH$_3$CN ratios in the four sources investigated by \citet{2023ApJS..267....4T} are $\approx0.001-0.04$.
Thus, the ratios at G28.28 are higher by factors of $\sim 2-70$, suggesting that G28.28 is the most cyanopolyyne-rich MYSO.

The CH$_3$OH/CH$_3$CN ratio at G28.28 is 24.7, whereas the ratios in the MYSOs studied by \citet{2023ApJS..267....4T} are 7 -- 1300.
Note that \citet{2023ApJS..267....4T} derived the column density of CH$_3$OH using its vibrational-excited line with a high $E_{\rm {up}}$ (302.9 K) with an assumption that its excitation temperatures are equal to those of CH$_3$CN.
This CH$_3$OH line traces only inner hot core regions.
On the other hand, the CH$_3$OH lines in the present work are likely contaminated by outer envelopes (Section \ref{sec:dis2}). 
Thus, the comparisons of the CH$_3$OH/CH$_3$CN ratio may be misleading.
If the CH$_3$OH column densities derived by \citet{2023ApJS..267....4T} trace only hot-core components, those values would be lower limits to the total CH$_3$OH column densities.
If the outer components of CH$_3$OH which are similar to those traced in our observations are included, the column densities of CH$_3$OH could be higher.
Thus, it can be said that CH$_3$OH is deficient at G28.28, especially in the hot region ($>100$ K), compared to COM-rich hot cores.

Here, we highlight that large COMs are especially deficient in G28.28. 
We could not detect CH$_3$CH$_2$CN ($12_{2,10}-11_{2,9}$; $E_{\rm {up}}=38.2$ K) and CH$_3$OCHO ($8_{3,6}-7_{3,5}$; $E_{\rm {up}}=27.3$ K) lines.
The previously studied MYSOs associated with HC$_5$N were found to be COM-rich; both nitrogen (N)-bearing and oxygen (O)-bearing COMs are abundant.
This is clearly different from G28.28.
Quantitative investigation with their upper limits will be discussed in Section \ref{sec:dis5}.

In summary, G28.28 has unique chemical features; cyanopolyynes are abundant, whereas COMs are deficient.
These features are the same as the low-mass WCCC source L1527 \citep{2013ChRv..113.8981S}.

\subsection{The Unique Chemical Features of the MYSO G28.28} \label{sec:dis5}

\citet{2023ApJS..267....4T} mentioned that HC$_5$N and large N-bearing COMs (CH$_2$CHCN and CH$_3$CH$_2$CN) are always detected together.
On the other hand, we could not detect large COMs (CH$_3$OCHO and CH$_3$CH$_2$CN) from G28.28, as we discussed in Section \ref{sec:dis3}. 
We derived upper limits for their column densities; $3.0\times10^{14}$ cm$^{-2}$ for CH$_3$OCHO and $1.2\times10^{13}$ cm$^{-2}$ for CH$_3$CH$_2$CN, respectively.
These correspond to fractional abundances with respect to H$_2$ of $2.5\times10^{-10}$ and $1.0\times10^{-11}$, respectively.
These abundances are lower than those in other hot cores by more than one order of magnitude \citep{2022MNRAS.512.4419P,2024MNRAS.533.1583L}.

\begin{figure}[ht!]
 \centering
  \includegraphics[bb = 40 10 240 245, scale = 0.82]{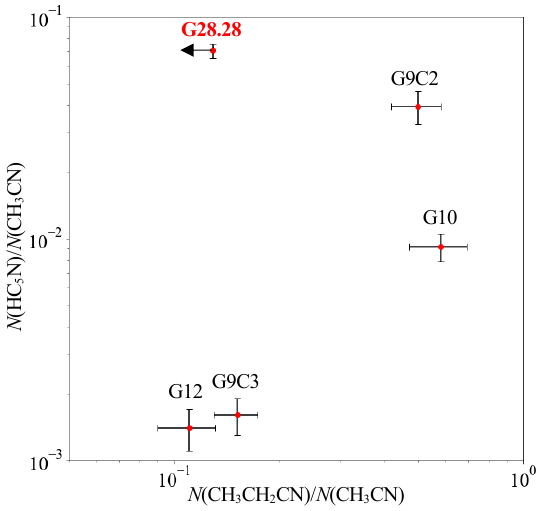} 
\caption{Comparison of column density ratios of HC$_5$N/CH$_3$CN vs. CH$_3$CH$_2$CN/CH$_3$CN among G28.28 and the previously studied MYSOs by \citet{2023ApJS..267....4T}. The error bars indicate the standard deviation.
\label{fig:MYSO}}
\end{figure}

The upper limit of the column density ratio of CH$_3$CH$_2$CN/CH$_3$CN at G28.28 is 0.13. 
The MYSOs studied by \citet{2023ApJS..267....4T} show the CH$_3$CH$_2$CN/CH$_3$CN column density ratios of 0.11 -- 0.58.
Figure \ref{fig:MYSO} shows comparisons of the column density ratios of HC$_5$N/CH$_3$CN vs. CH$_3$CH$_2$CN/CH$_3$CN among G28.28 and the MYSOs studied by \citet{2023ApJS..267....4T}.
The previously studied MYSOs are divided into two groups: both ratios are high (G9C2 and G10) or low (G9C3 and G12).
G28.28 is distinct from these two groups, showing a higher HC$_5$N/CH$_3$CN ratio in comparison to the other MYSOs.
On the other hand, the upper limit of the CH$_3$CH$_2$CN/CH$_3$CN ratio is lower than those in G9C2 and G10 by a factor of $\sim 4-4.5$.
G12 and G9C3 show similar CH$_3$CH$_2$CN/CH$_3$CN ratios with the upper limit at G28.28, but their HC$_5$N/CH$_3$CN ratios are lower than that at G28.28 by a factor of $\sim 43-50$. 
Even if we consider the different sensitivities and beam sizes among our observations and the previous study, the differences in the column density ratios would be less than one order of magnitude.
Thus, G28.28 is genuinely deficient in COMs, making its chemical composition different from those of the other MYSOs studied by \citet{2023ApJS..267....4T}.

Although G28.28 contains hot gas with temperatures above 100 K, where ice mantles sublimate, COMs are significantly deficient. 
This indicates that the COM-poor nature of G28.28 is not simply due to low temperature.
In our ALMA observations, the frequency coverage is limited.
We then checked the previous single-dish observations covering a wide frequency range \citep{2018ApJ...866..150T}. 
They detected only two COMs, CH$_3$OH and CH$_3$CHO, whereas many lines of COMs have been detected from the other two MYSOs \citep[see Table 6 of][]{2018ApJ...866..150T}.

In summary, the unique chemical features of G28.28 are (1) the HCCC mechanism produces HC$_5$N in the hot region with temperatures above 100 K, and (2) COMs are deficient even though G28.28 harbors a hot region.
To the best of our knowledge, G28.28 is the first HCCC source, in the categorization of the chemical diversity around protostars proposed by \citet{2024ApSS.369...34T}.

\section{Conclusion}\label{sec:con}

We have presented ALMA Band 3 data toward the MYSO G28.28.
Both cyanopolyynes (HC$_3$N and HC$_5$N) and COMs (CH$_3$OH, CH$_3$CN, and CH$_3$CHO) have been detected.
On the other hand, large COMs (CH$_3$CH$_2$CN and CH$_3$OCHO) were not detected.
The main conclusions of this paper are as follows: 

\begin{enumerate}
    \item The detected HC$_5$N and CH$_3$CN lines from G28.28 trace a hot core region with temperatures above 100 K. These results indicate that HC$_5$N forms by the HCCC mechanism. 
    \item Additional HC$_5$N peak positions have been identified between the MYSO G28.28 and the nearby UC\ion{H}{2} region. Their densities and excitation temperatures are consistent with dense, moderately warm gas, possibly associated with low- or intermediate-mass protostellar cores. The chemistry at these positions may be related to WCCC-like carbon-chain formation, although external heating from the nearby UC\ion{H}{2} region cannot be excluded with the present data.
    \item The high-sensitivity, high-angular resolution ALMA Band 3 data confirm that G28.28 is cyanopolyyne-rich but COM-poor.
    Such a chemical feature is different from the MYSOs previously studied with ALMA Band 3 \citep{2023ApJS..267....4T}.
\end{enumerate}

We found that G28.28 is the counterpart of the low-mass WCCC source L1527.
The MYSO G28.28 is the first HCCC source; a cyanopolyyne-rich but COM-poor high-mass protostellar core harboring a hot region ($T>100$ K).

Understanding the origin(s) of this chemical diversity will be key to improving our understanding of the star-formation process, including the role of natal environments.
These points will be studied with large samples located in various environments. 

Most chemical network simulations show that the most efficient formation route of cyanopolyynes is the reactions of ``C$_{2n}$H$_2$ + CN'' during the warm-up stage \citep{2009MNRAS.394..221C, 2019ApJ...881...57T}, and these reactions are proposed as the most promising formation pathway of cyanopolyynes in the HCCC mechanism at present. However, it is difficult to directly prove it by observations. One method to investigate the formation process of long cyanopolyynes is observations of $^{13}$C isotopologues \citep{2016ApJ...817..147T, 2018MNRAS.474.5068B}. Such observations are currently difficult even with ALMA due to required long integration times, because we need to detect the $^{13}$C isotopologues with S/N $>10$. Although the concentration of $^{13}$C in CN may not be efficient in the warm gas, \citet{2016ApJ...830..106T} demonstrated that this method can be applied to HC$_3$N in the warm gas around both low-mass and high-mass protostars. Observations with future facilities are necessary for direct evidence of the HCCC mechanism.

%% Please use the acknowledgment and contribution environments. This will 
%% be anonomyized when the "anonymous" style option is used. 
\begin{acknowledgments}
This paper makes use of the following ALMA data: ADS/JAO. ALMA\#2023.1.00467.S. ALMA is a partnership of ESO (representing its member states), NSF (USA) and NINS (Japan), together with NRC (Canada), NSTC and ASIAA (Taiwan), and KASI (Republic of Korea), in cooperation with the Republic of Chile. The Joint ALMA Observatory is operated by ESO, AUI/NRAO and NAOJ.
Data analysis was in part carried out on the Multi-wavelength Data Analysis System operated by the Astronomy Data Center (ADC), National Astronomical Observatory of Japan.

K.T. is supported by JSPS KAKENHI grant Nos. 21H01142, 24K17096, and 24H00252.
O.S. acknowledges the financial support of the Max Planck Society.
T.S. acknowledges the financial support of the Yamada Science Foundation.
We also thank the anonymous referee for the valuable feedback that has improved the quality of this work.
\end{acknowledgments}

%\begin{contribution}
%%This section gives authors the space to recognize author contributions. The text inside this environment is NOT counted towards the total word quanta. At a minimum, manuscripts are expected to include this text:

%All authors contributed equally to the Terra Mater collaboration.

%% But authors are expected to provide more specific details, e.g. 
%%
%%SC was responsible for writing and submitting the manuscript.
%%WWM came up with the initial research concept and edited the manuscript.
%%OTS obtained the funding and edited the manuscript.
%%EBF provided the formal analysis and validation. He also edited the manuscript.
%%GEH Supervised the undergraduates, wrote the software and administers the project github and Zenodo repositories.
%%
%% Authors can use the Contributor Role Taxonomy (CRediT) at
%% for ideas on how write a good statement tailored to their needs.

%\end{contribution}

\facilities{Atacama Large Millimeter/submillimeter Array (ALMA)}

\software{CASA \citep{2022PASP..134k4501C},  
          CASSIS \citep{2015sf2a.conf..313V}, 
          }


\appendix

\section{SED fitting of Infrared Data} \label{sec:sed}

\begin{figure}[ht!]
 \centering
  \includegraphics[bb = 150 20 280 280, scale = 0.6]{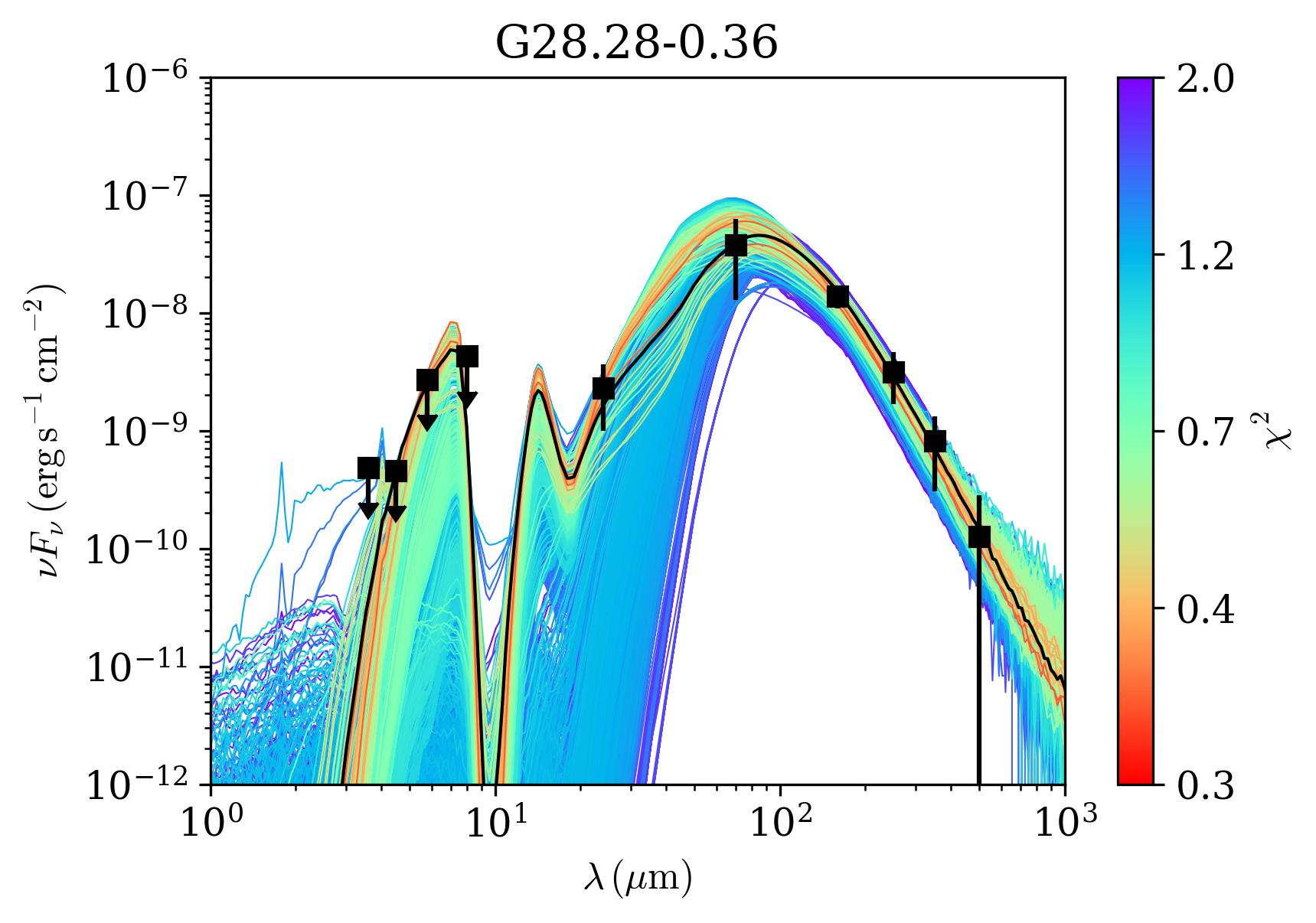} 
\caption{SED fitting of the MYSO G28.28. The fixed-aperture, background-subtracted SED was fitted using the protostellar model grid \citep{2018ApJ...853...18Z}. The best-fitting model is shown by the black line, while all other ``good models" are displayed by colored lines, ranging from red to blue with increasing $\chi^2$. We define ``good'' models depending on the minimum value of $\chi^2$, $i.e.$, $\chi^2_{min}$ \citep[see][]{2023ApJ...942....7F,2025ApJ...986...15T}.
\label{fig:sed}}
\end{figure}

\begin{figure*}[ht!]
 \centering
  \includegraphics[bb =60 10 680 200, scale = 0.7]{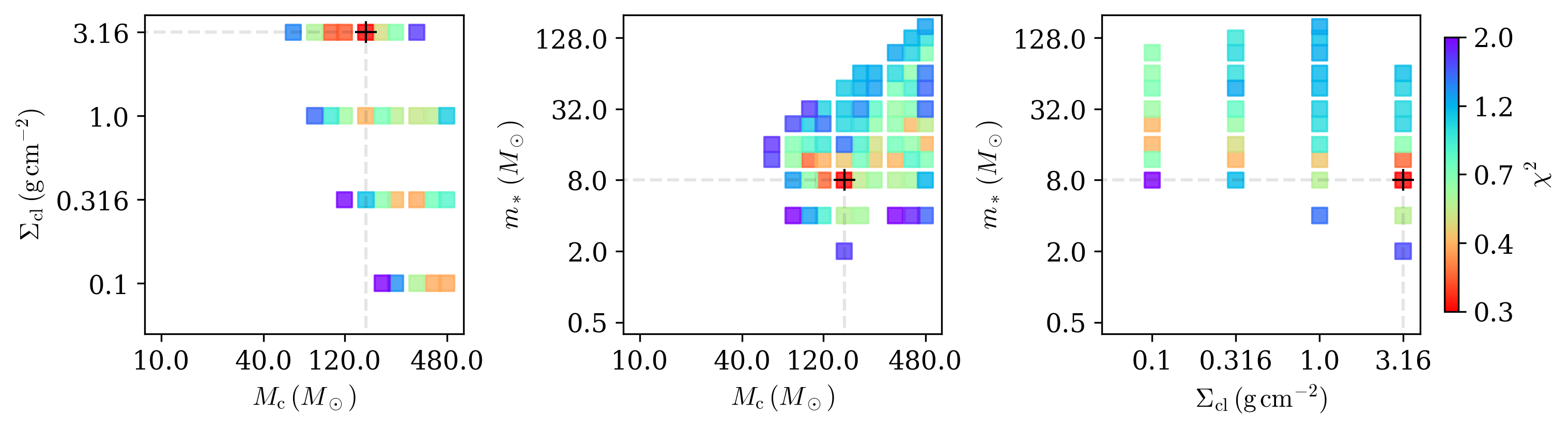} 
\caption{Comparison of key model parameters for source G28.28-0.36. The panels show $\Sigma_{\mathrm{cl}}$ versus $M_{\mathrm{c}}$ (left column), $m_{*}$ versus $M_{\mathrm{c}}$ (center column), and $m_{*}$ versus $\Sigma_{\mathrm{cl}}$ (right column) for ``good'' model fits, color-coded by their $\chi^2$ values. The black cross indicates the best-fitting model.
\label{fig:sed-para}}
\end{figure*}

For the spectral energy distribution (SED) fitting, we used archival data from \textit{Spitzer}/IRAC \citep{2004ApJS..154....1W, 2004ApJS..154...10F} at 3.6, 4.5, 5.8, and 8.0~$\mu$m obtained from the \textit{Spitzer} Heritage Archive, along with \textit{Herschel}/PACS and SPIRE \citep{2010A&A...518L...3G} data at 70, 160, 250, 350, and 500~$\mu$m from the ESA \textit{Herschel} Science Archive. In addition, \textit{Spitzer}/MIPS 24~$\mu$m data were included. The \textit{Herschel} 70~$\mu$m image was used to automatically determine the optimal aperture using specialized functions from \texttt{sedcreator} \citep{2023ApJ...942....7F}, yielding an aperture of 25\arcsec (radius). 
This aperture was then adopted as a fixed aperture to extract fluxes at all wavelengths. Background subtraction was performed using an annulus with an inner radius equal to the aperture radius and an outer radius of twice the aperture radius. Figure \ref{fig:sed} presents the resulting SED fit, and Figure \ref{fig:sed-para} shows the corresponding two-dimensional parameter space of the model fits. The best-fit SED parameters are summarized in Table~\ref{tab:sedfit}. The reported model parameters and derived quantities include the $\chi^2$ goodness-of-fit statistic; $M_c$, the core mass; $\Sigma_{\rm cl}$, the mass surface density of the surrounding clump; $R_c$, the core radius; $m_\ast$, the stellar mass; $\theta_{\rm view}$, the viewing angle measured from the outflow axis; $A_V$, the line-of-sight visual extinction; $M_{\rm env}$, the envelope mass; $\theta_{\rm w,esc}$, the half-opening angle of the outflow cavity (escape angle); $\dot{M}_{\rm disk}$, the disk accretion rate onto the star; $L_{\rm bol,iso}$, the isotropic bolometric luminosity; and $L_{\rm bol}$, the true bolometric luminosity of the source.

\begin{deluxetable}{cccccccccccc}
\tablewidth{0pt} 
\tablecaption{Selected SED model parameters \label{tab:sedfit}}
\tablehead{\colhead{$\chi^2$} & \colhead{$M_c$} & \colhead{$\Sigma_{\mathrm{cl}}$} & \colhead{$R_c$} & \colhead{$m_*$} & \colhead{$\theta_{\mathrm{view}}$} & \colhead{$A_V$} & \colhead{$M_{\mathrm{env}}$} & \colhead{$\theta_{\mathrm{w,esc}}$} & \colhead{$\dot{M}_{\mathrm{disk}}$} & \colhead{ $L_{\mathrm{bol,iso}}$} & \colhead{$L_{\mathrm{bol}}$} \\
\colhead{} & \colhead{($M_\odot$)} & \colhead{(g\,cm$^{-2}$)} & \colhead{(pc)} & \colhead{($M_\odot$)} & \colhead{(deg)} & \colhead{(mag)} & \colhead{($M_\odot$)} & \colhead{(deg)} & \colhead{($\times 10^{-4}$ $M_\odot\,\mathrm{yr}^{-1}$)} & \colhead{($\times 10^4$ $L_\odot$)} & \colhead{($\times 10^4$ $L_\odot$)}
}
\startdata
0.18510 & 160.0 & 3.16 & 0.052362 & 8.0 & 12.83857 & 201.35329 & 145.8153 & 12.7017 & 8.51410 & 9.70607 & 2.00533 \\
0.26705 & 120.0 & 3.16 & 0.045347 & 8.0 & 12.83857 & 214.48997 & 105.9249 & 14.9547 & 7.89885 & 11.8794 & 1.89690 \\
0.23528 & 100.0 & 3.16 & 0.041396 & 12.0 & 12.83857 & 279.24634 & 76.72601 & 20.2152 & 9.39717 & 41.1093 & 5.24975 \\
0.28694 & 120.0 & 3.16 & 0.045347 & 12.0 & 12.83857 & 281.13601 & 98.53957 & 17.8019 & 9.63771 & 41.7014 & 5.17369 \\
0.25148 & 480.0 & 0.1 & 0.50982 & 16.0 & 12.83857 & 234.00543 & 440.0145 & 15.4996 & 1.19434 & 18.6013 & 3.85823 \\
Average model\tablenotemark{a} & $196^{+284}_{-96}$	& $2.5^{+0.6}_{-2.4}$ & $0.14^{+0.37}_{-0.10}$ & $11.2^{+4.8}_{-3.2}$ &	12.83857 & $242^{+39}_{-41}$ &	$173^{+267}_{-97}$ & $16 \pm 4$ & $7.3^{+2.3}_{-6.1}$ & $24.6^{+17.1}_{-14.9}$ & $3.6^{+1.6}_{-1.7}$ \\
\enddata
\tablenotetext{a}{This corresponds to the average values of the above five good models. Upper and lower scripts in the average model row refer to the upper and lower range of the dispersion of the good models.}
\end{deluxetable}

\section{Calculation of $N$(H$_2$)} \label{sec:append1}

We calculated the dust opacity at G28.28 and Peak\,1 -- Peak\,4 using the following Equation (\ref{equ:taucont}):
\begin{equation} \label{equ:taucont}
\tau = -{\rm{ln}}\Big(1-\frac{F_{2.75 {\rm {mm}}} }{B_{2.75{\rm {mm}}}(T_{\rm {dust}}) \Omega}\Big), 
\end{equation}
where $F_{2.75 {\rm {mm}}}$, $B_{2.75{\rm {mm}}}$($T_{\rm {dust}}$), $\Omega$ are flux density integrated over the apertures listed in Table \ref{tab:HC5Npeak}, the Planck function at 2.75 mm with a dust temperature of $T_{\rm {dust}}$, and the solid angle of the aperture, respectively.
We assumed that the dust temperature is equal to the excitation temperature of CH$_3$CN \citep{2023ApJ...950...57T}.
Since we applied two-velocity component fitting to the MYSO G28.28, we used their average temperature (75.4 K) as the dust temperature. 
The obtained dust opacity values are 0.001 -- 0.007 at all positions.
Thus, the dust continuum emission at this frequency is optically thin in our target positions.

We derived the column density of H$_2$, $N$(H$_2$), from the dust continuum data at 2.75 mm using the following Equation (\ref{equ:NH2}):

\begin{equation} \label{equ:NH2}
N({\rm {H}}_{2}) = \frac{F_{2.75 {\rm {mm}}} \gamma}{B_{2.75{\rm {mm}} }(T_{\rm {dust}}) \Omega \kappa_{2.75{\rm {mm}} }  \mu_{\rm {H}_2} m_{\rm {H}}}, 
\end{equation}
where $F_{2.75 {\rm {mm}}}$, $\gamma$, $B_{2.75{\rm {mm}}}$($T_{\rm {dust}}$), $\Omega$, $\kappa_{2.75{\rm {mm}}}$, $\mu_{\rm {H}_2}$, and $m_{\rm {H}}$ are continuum flux density at 2.75 mm, gas-to-dust mass ratio, the Planck function at 2.75 mm with a dust temperature of $T_{\rm {dust}}$, the solid angle of the aperture, the dust mass opacity, the H$_{2}$ mean molecular weight (2.8), and the mass of the hydrogen atom, respectively.

The gas-to-dust ratio was derived using the following Equation (\ref{equ:gtd}) \citep{2017AA...606L..12G}:
\begin{equation} \label{equ:gtd}
{\rm {log}}(\gamma) = 0.087 D_{\rm {GC}} + 1.44,
\end{equation}
where $D_{\rm {GC}}$ represents the distance from the Galactic Center.
The $D_{\rm {GC}}$ value for G28.28 was calculated at 5.5 kpc \citep{2021ApJ...908..100T}.
The calculated dust-to-gas ratio is 83.

The $\kappa_{2.75{\rm {mm}}}$ value was calculated with the following Equation (\ref{equ:kappa}):
\begin{equation} \label{equ:kappa}
\kappa_{2.75{\rm {mm}}} = \kappa_{1.3{\rm {mm}}} \Big(\frac{\nu_{2.75{\rm {mm}}}}{\nu_{1.3{\rm {mm}}}}\Big)^\beta, 
\end{equation} \\
where $\kappa_{1.3{\rm {mm}}}$ is the dust mass opacity at 1.3\,mm and $\beta=\alpha-2$ \citep{1993ApJ...414..759T}.
The $\kappa_{1.3{\rm {mm}}}$ value of 0.9 cm$^2$\,g$^{-1}$ \citep{1994AA...291..943O,2023ApJ...950...57T} was applied.
We assumed that the dust emissivity ($\beta$) is equal to 1.5 here.
\citet{2023ApJS..267....4T} derived the $\beta$ values of 1.5 -- 1.75 at the five MYSOs with similar ages using multi-band infrared data.
In addition, \citet{2024AA...687A.163B} derived the H$_2$ column densities for hot cores with an assumption of $\beta = 1.5$.
Thus, we considered that $\beta = 1.5$ can be applied to G28.28.
The H$_2$ density values, $n$(H$_2$), summarized in Table \ref{tab:HC5Npeak} were calculated with the assumption that the size along the line of sight of the core is equal to the aperture sizes listed in Table \ref{tab:HC5Npeak}. This assumption is reasonable for G28.28, showing a centrally peaked feature. On the other hand, this assumption may not be applicable to Peak\,1 -- Peak\,4, because there is no centrally peaked feature and hence the determination of the appropriate line-of-sight scale is not straightforward. If the line-of-sight scale is underestimated, the derived $n$(H$_2$) would be upper limits. Even if the line-of-sight scale is several factors larger than assumed here, the H$_2$ density is still higher than $10^5$ cm$^{-3}$. Thus, our assumption of the LTE condition in the molecular line analyses is reasonable.

\section{Spectra at the four HC$_5$N peaks} \label{sec:append2}

\begin{figure*}[ht!]
 \centering
  \includegraphics[bb = 0 0 530 500, scale = 0.9]{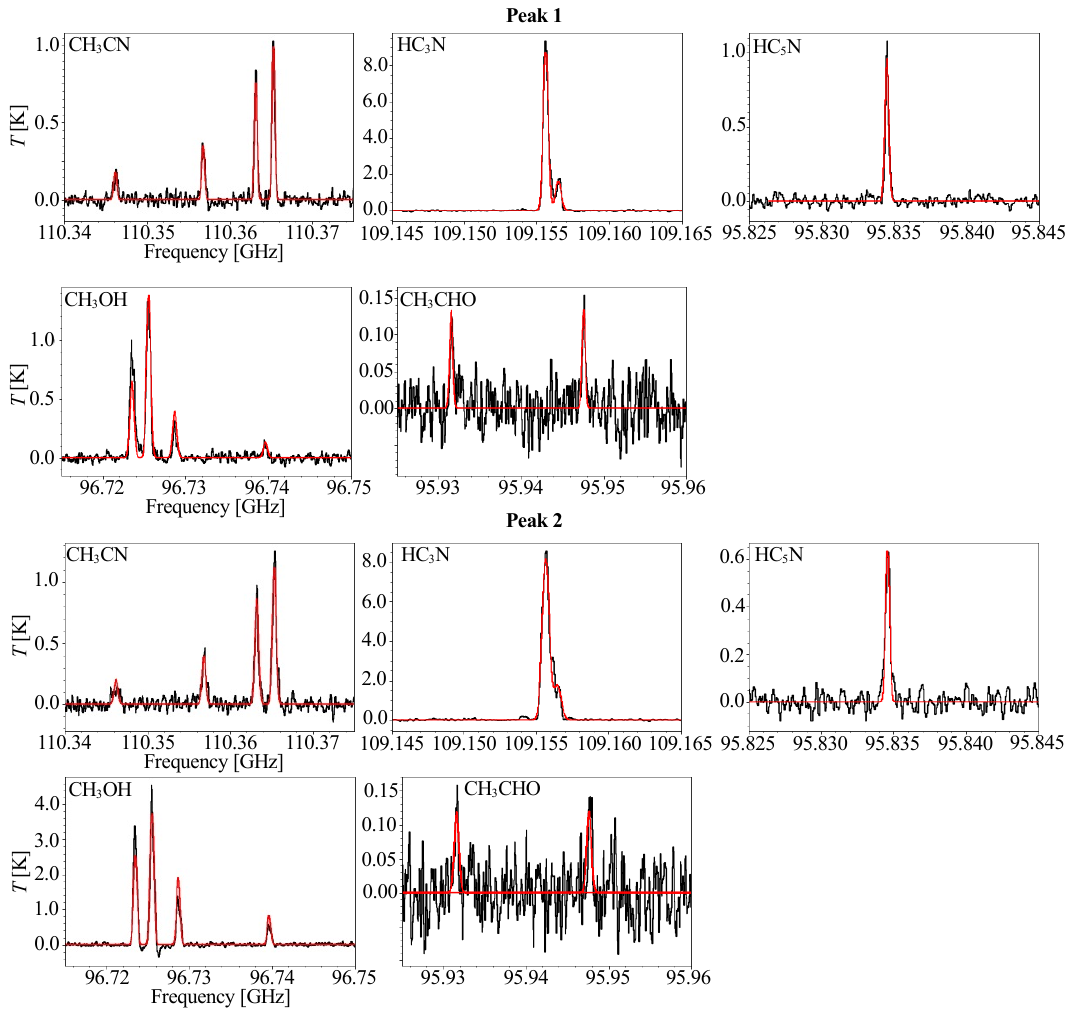} 
\caption{Spectra at Peak\,1 (upper panels) and Peak\,2 (lower panels). Black and red lines indicate the observed spectra and the best-fitting models by CASSIS, respectively. \label{fig:specP12}}
\end{figure*}

\begin{figure*}[ht!]
 \centering
  \includegraphics[bb = 0 0 530 500, scale = 0.9]{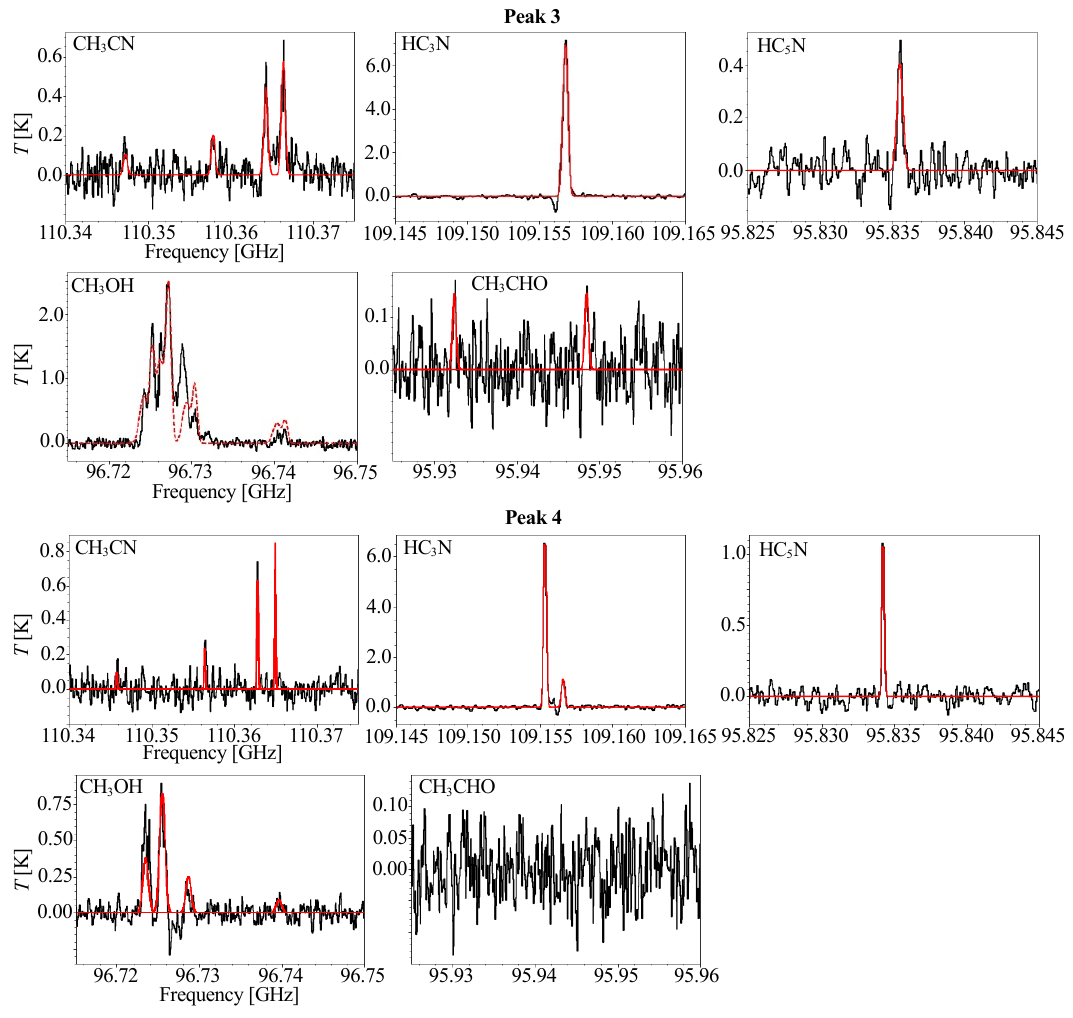} 
\caption{Spectra at Peak\,3 (upper panels) and Peak\,4 (lower panels). Black and red lines indicate the observed spectra and the best-fitting models by CASSIS, respectively. \label{fig:specP34}}
\end{figure*}

Table \ref{tab:line} summarizes line parameters used in the MCMC fitting.
Figures \ref{fig:specP12} and \ref{fig:specP34} show spectra at the four HC$_5$N peak positions (Peak\,1 -- Peak\,4), overlaid with the best-fitting models derived in the CASSIS software. 
The CH$_3$OH lines at Peak\,3 could not be fitted even with two velocity components. 
The fitting results are tentative and shown by the dotted lines. 
No CH$_3$CHO line was detected at Peak\,4.

\begin{deluxetable}{llcc}[h]
\tablewidth{0pt}
\tablecaption{Line parameters used for the MCMC fitting \label{tab:line}}
\tablehead{\colhead{Species} & \colhead{Transition} & \colhead{Frequency (GHz)\tablenotemark{a}} & \colhead{$E_{\rm {up}}$ (K)}
}
\startdata
      CH$_3$CN & $6_3-5_3$ & 110.36435 & 82.8 \\
      CH$_3$CN & $6_2-5_2$ & 110.37499 & 47.1 \\
      CH$_3$CN & $6_1-5_1$ & 110.38137 & 25.7 \\
      CH$_3$CN & $6_0-5_0$ & 110.38350 & 18.5 \\
      HC$_3$N  & $12-11$   & 109.17363 & 34.1 \\
      HC$_5$N  & $36-35$   & 95.850335 & 85.1 \\
      CH$_3$OH & $2_{1,2}-1_{1,1}$ $E$ & 96.739358 & 12.5 \\
      CH$_3$OH & $2_{0,2}-1_{0,1}$ $A$ & 96.741371 & 7.0 \\
      CH$_3$OH & $2_{-0,2}-1_{-0,1}$ $E$ & 96.744545 & 20.1 \\
      CH$_3$OH & $2_{-1,1}-1_{-1,0}$ $E$ & 96.755501 & 28.0 \\
      CH$_3$CHO & $5_{0,5}-4_{0,4}$ $E$ & 95.9474373 & 13.9 \\
      CH$_3$CHO & $5_{0,5}-4_{0,4}$ $A$ & 95.9634588 & 13.9 \\
\enddata
\tablenotetext{a}{Rest frequencies are taken from the Cologne Database for Molecular Spectroscopy (CDMS; \cite{2016JMoSp.327...95E}) except for CH$_3$CHO, whose frequencies are taken from the JPL catalog \citep{1998JQSRT..60..883P}.}
\end{deluxetable}

%% For this sample we use BibTeX plus aasjournalv7.bst to generate the
\clearpage
\bibliography{ALMA_Band3_G28}{}

@ARTICLE{2024AA...687A.163B,
       author = {{Bonfand}, M. and {Csengeri}, T. and {Bontemps}, S. and {Brouillet}, N. and {Motte}, F. and {Louvet}, F. and {Ginsburg}, A. and {Cunningham}, N. and {Galv{\'a}n-Madrid}, R. and {Herpin}, F. and {Wyrowski}, F. and {Valeille-Manet}, M. and {Stutz}, A.~M. and {Di Francesco}, J. and {Gusdorf}, A. and {Fern{\'a}ndez-L{\'o}pez}, M. and {Lefloch}, B. and {Liu}, H.-L. and {Sanhueza}, P. and {{\'A}lvarez-Guti{\'e}rrez}, R.~H. and {Olguin}, F. and {Nony}, T. and {Lopez-Sepulcre}, A. and {Dell'Ova}, P. and {Pouteau}, Y. and {Jeff}, D. and {Chen}, H.-R.~V. and {Armante}, M. and {Towner}, A. and {Bronfman}, L. and {Kessler}, N.},
        title = "{ALMA-IMF. XI. The sample of hot core candidates: A rich population of young high-mass protostars unveiled by the emission of methyl formate}",
      journal = {\aap},
         year = 2024,
        month = jul,
       volume = {687},
          eid = {A163},
        pages = {A163},
          doi = {10.1051/0004-6361/202347856},
archivePrefix = {arXiv},
       eprint = {2402.15023},
 primaryClass = {astro-ph.GA},
       adsurl = {https://ui.adsabs.harvard.edu/abs/2024A&A...687A.163B}
}

@ARTICLE{2024ApSS.369...34T,
       author = {{Taniguchi}, Kotomi and {Gorai}, Prasanta and {Tan}, Jonathan C.},
        title = "{Carbon-chain chemistry in the interstellar medium}",
      journal = {\apss},
         year = 2024,
        month = apr,
       volume = {369},
       number = {4},
          eid = {34},
        pages = {34},
          doi = {10.1007/s10509-024-04292-9},
archivePrefix = {arXiv},
       eprint = {2303.15769},
 primaryClass = {astro-ph.GA},
       adsurl = {https://ui.adsabs.harvard.edu/abs/2024Ap&SS.369...34T}
}

@ARTICLE{2024AA...692A..65T,
       author = {{Taniguchi}, Kotomi and {Gorai}, Prasanta and {Tan}, Jonathan C. and {G{\'o}mez-Garrido}, Miguel and {Fedriani}, Rub{\'e}n and {Yang}, Yao-Lun and {Tirupati Kumara}, Sridharan and {Tanaka}, Kei E.~I. and {Saito}, Masao and {Zhang}, Yichen and {Morgan}, Lawrence and {Cosentino}, Giuliana and {Law}, Chi-Yan},
        title = "{The SOFIA Massive (SOMA) Star Formation Q-band follow-up: I. Carbon-chain chemistry of intermediate-mass protostars}",
      journal = {\aap},
         year = 2024,
        month = dec,
       volume = {692},
          eid = {A65},
        pages = {A65},
          doi = {10.1051/0004-6361/202451499},
archivePrefix = {arXiv},
       eprint = {2410.23392},
 primaryClass = {astro-ph.GA},
       adsurl = {https://ui.adsabs.harvard.edu/abs/2024A&A...692A..65T}
}

@ARTICLE{2019ApJ...881...57T,
       author = {{Taniguchi}, Kotomi and {Herbst}, Eric and {Caselli}, Paola and {Paulive}, Alec and {Maffucci}, Dominique M. and {Saito}, Masao},
        title = "{Cyanopolyyne Chemistry around Massive Young Stellar Objects}",
      journal = {\apj},
         year = 2019,
        month = aug,
       volume = {881},
       number = {1},
          eid = {57},
        pages = {57},
          doi = {10.3847/1538-4357/ab2d9e},
archivePrefix = {arXiv},
       eprint = {1906.11296},
 primaryClass = {astro-ph.GA},
       adsurl = {https://ui.adsabs.harvard.edu/abs/2019ApJ...881...57T}
}

@ARTICLE{2021ApJ...908..100T,
       author = {{Taniguchi}, Kotomi and {Herbst}, Eric and {Majumdar}, Liton and {Caselli}, Paola and {Tan}, Jonathan C. and {Li}, Zhi-Yun and {Shimoikura}, Tomomi and {Dobashi}, Kazuhito and {Nakamura}, Fumitaka and {Saito}, Masao},
        title = "{Carbon Chain Chemistry in Hot-core Regions around Three Massive Young Stellar Objects Associated with 6.7 GHz Methanol Masers}",
      journal = {\apj},
         year = 2021,
        month = feb,
       volume = {908},
       number = {1},
          eid = {100},
        pages = {100},
          doi = {10.3847/1538-4357/abd6c9},
archivePrefix = {arXiv},
       eprint = {2012.12993},
 primaryClass = {astro-ph.GA},
       adsurl = {https://ui.adsabs.harvard.edu/abs/2021ApJ...908..100T}
}

@ARTICLE{2023ApJS..267....4T,
       author = {{Taniguchi}, Kotomi and {Majumdar}, Liton and {Caselli}, Paola and {Takakuwa}, Shigehisa and {Hsieh}, Tien-Hao and {Saito}, Masao and {Li}, Zhi-Yun and {Dobashi}, Kazuhito and {Shimoikura}, Tomomi and {Nakamura}, Fumitaka and {Tan}, Jonathan C. and {Herbst}, Eric},
        title = "{Chemical Differentiation around Five Massive Protostars Revealed by ALMA: Carbon-chain Species and Oxygen/Nitrogen-bearing Complex Organic Molecules}",
      journal = {\apjs},
         year = 2023,
        month = jul,
       volume = {267},
       number = {1},
          eid = {4},
        pages = {4},
          doi = {10.3847/1538-4365/acd110},
archivePrefix = {arXiv},
       eprint = {2304.13873},
 primaryClass = {astro-ph.GA},
       adsurl = {https://ui.adsabs.harvard.edu/abs/2023ApJS..267....4T}
}

@ARTICLE{2018ApJ...866...32T,
       author = {{Taniguchi}, Kotomi and {Miyamoto}, Yusuke and {Saito}, Masao and {Sanhueza}, Patricio and {Shimoikura}, Tomomi and {Dobashi}, Kazuhito and {Nakamura}, Fumitaka and {Ozeki}, Hiroyuki},
        title = "{Interferometric Observations of Cyanopolyynes toward the G28.28-0.36 High-mass Star-forming Region}",
      journal = {\apj},
         year = 2018,
        month = oct,
       volume = {866},
       number = {1},
          eid = {32},
        pages = {32},
          doi = {10.3847/1538-4357/aadd0c},
archivePrefix = {arXiv},
       eprint = {1808.08435},
 primaryClass = {astro-ph.SR},
       adsurl = {https://ui.adsabs.harvard.edu/abs/2018ApJ...866...32T}
}

@ARTICLE{2023ApJ...950...57T,
       author = {{Taniguchi}, Kotomi and {Sanhueza}, Patricio and {Olguin}, Fernando A. and {Gorai}, Prasanta and {Das}, Ankan and {Nakamura}, Fumitaka and {Saito}, Masao and {Zhang}, Qizhou and {Lu}, Xing and {Li}, Shanghuo and {Chen}, Huei-Ru Vivien},
        title = "{Digging into the Interior of Hot Cores with the ALMA (DIHCA). III. The Chemical Link between NH$_{2}$CHO, HNCO, and H$_{2}$CO}",
      journal = {\apj},
         year = 2023,
        month = jun,
       volume = {950},
       number = {1},
          eid = {57},
        pages = {57},
          doi = {10.3847/1538-4357/acca1d},
archivePrefix = {arXiv},
       eprint = {2304.00267},
 primaryClass = {astro-ph.GA},
       adsurl = {https://ui.adsabs.harvard.edu/abs/2023ApJ...950...57T}
}

@INPROCEEDINGS{2015sf2a.conf..313V,
       author = {{Vastel}, C. and {Bottinelli}, S. and {Caux}, E. and {Glorian}, J.-M. and {Boiziot}, M.},
        title = "{CASSIS: a tool to visualize and analyse instrumental and synthetic spectra.}",
    booktitle = {SF2A-2015: Proceedings of the Annual meeting of the French Society of Astronomy and Astrophysics},
         year = 2015,
       editor = {{Martins}, F. and {Boissier}, S. and {Buat}, V. and {Cambr{\'e}sy}, L. and {Petit}, P.},
        month = dec,
        pages = {313-316},
       adsurl = {https://ui.adsabs.harvard.edu/abs/2015sf2a.conf..313V}
}

@ARTICLE{1993ApJ...414..759T,
       author = {{Terebey}, S. and {Chandler}, C.~J. and {Andre}, P.},
        title = "{The Contribution of Disks and Envelopes to the Millimeter Continuum Emission from Very Young Low-Mass Stars}",
      journal = {\apj},
         year = 1993,
        month = sep,
       volume = {414},
        pages = {759},
          doi = {10.1086/173121},
       adsurl = {https://ui.adsabs.harvard.edu/abs/1993ApJ...414..759T}
}

@ARTICLE{2013ChRv..113.8981S,
       author = {{Sakai}, Nami and {Yamamoto}, Satoshi},
        title = "{Warm Carbon-Chain Chemistry}",
      journal = {Chemical Reviews},
         year = 2013,
        month = dec,
       volume = {113},
       number = {12},
        pages = {8981-9015},
          doi = {10.1021/cr4001308},
       adsurl = {https://ui.adsabs.harvard.edu/abs/2013ChRv..113.8981S}
}

@ARTICLE{1998JQSRT..60..883P,
       author = {{Pickett}, H.~M. and {Poynter}, R.~L. and {Cohen}, E.~A. and {Delitsky}, M.~L. and {Pearson}, J.~C. and {M{\"u}ller}, H.~S.~P.},
        title = "{Submillimeter, millimeter and microwave spectral line catalog.}",
      journal = {\jqsrt},
         year = 1998,
        month = nov,
       volume = {60},
       number = {5},
        pages = {883-890},
          doi = {10.1016/S0022-4073(98)00091-0},
       adsurl = {https://ui.adsabs.harvard.edu/abs/1998JQSRT..60..883P}
}

@ARTICLE{2014ApJ...788..187H,
       author = {{Hunter}, T.~R. and {Brogan}, C.~L. and {Cyganowski}, C.~J. and {Young}, K.~H.},
        title = "{Subarcsecond Imaging of the NGC 6334 I(N) Protocluster: Two Dozen Compact Sources and a Massive Disk Candidate}",
      journal = {\apj},
         year = 2014,
        month = jun,
       volume = {788},
       number = {2},
          eid = {187},
        pages = {187},
          doi = {10.1088/0004-637X/788/2/187},
archivePrefix = {arXiv},
       eprint = {1405.0496},
 primaryClass = {astro-ph.SR},
       adsurl = {https://ui.adsabs.harvard.edu/abs/2014ApJ...788..187H}
}

@ARTICLE{2025ApJ...987..197H,
       author = {{Hoque}, Ariful and {Baug}, Tapas and {Dewangan}, Lokesh K. and {Juvela}, Mika and {Tej}, Anandmayee and {Goldsmith}, Paul F. and {Garc{\'\i}a}, Pablo and {Stutz}, Amelia M. and {Liu}, Tie and {Lee}, Chang Won and {Xu}, Fengwei and {Sanhueza}, Patricio and {Bhadari}, N.~K. and {Tatematsu}, K. and {Liu}, Xunchuan and {Liu}, Hong-Li and {Zhang}, Yong and {Tang}, Xindi and {Garay}, Guido and {Wang}, Ke and {Zhang}, Siju and {T{\'o}th}, L. Viktor and {Nazeer}, Hafiz and {Hwang}, Jihye and {Gorai}, Prasanta and {Bronfman}, Leonardo and {Das}, Swagat Ranjan and {Sinha}, Tirthendu},
        title = "{The ALMA-ATOMS Survey: Exploring Protostellar Outflows in HC$_{3}$N}",
      journal = {\apj},
         year = 2025,
        month = jul,
       volume = {987},
       number = {2},
          eid = {197},
        pages = {197},
          doi = {10.3847/1538-4357/add928},
archivePrefix = {arXiv},
       eprint = {2505.04164},
 primaryClass = {astro-ph.GA},
       adsurl = {https://ui.adsabs.harvard.edu/abs/2025ApJ...987..197H}
}

@ARTICLE{2014ApJ...786...38H,
       author = {{Hern{\'a}ndez-Hern{\'a}ndez}, Vicente and {Zapata}, Luis and {Kurtz}, Stan and {Garay}, Guido},
        title = "{SMA Millimeter Observations of Hot Molecular Cores}",
      journal = {\apj},
         year = 2014,
        month = may,
       volume = {786},
       number = {1},
          eid = {38},
        pages = {38},
          doi = {10.1088/0004-637X/786/1/38},
archivePrefix = {arXiv},
       eprint = {1402.2682},
 primaryClass = {astro-ph.GA},
       adsurl = {https://ui.adsabs.harvard.edu/abs/2014ApJ...786...38H}
}

@ARTICLE{2017AA...606L..12G,
       author = {{Giannetti}, A. and {Leurini}, S. and {K{\"o}nig}, C. and {Urquhart}, J.~S. and {Pillai}, T. and {Brand}, J. and {Kauffmann}, J. and {Wyrowski}, F. and {Menten}, K.~M.},
        title = "{Galactocentric variation of the gas-to-dust ratio and its relation with metallicity}",
      journal = {\aap},
         year = 2017,
        month = oct,
       volume = {606},
          eid = {L12},
        pages = {L12},
          doi = {10.1051/0004-6361/201731728},
archivePrefix = {arXiv},
       eprint = {1710.05721},
 primaryClass = {astro-ph.GA},
       adsurl = {https://ui.adsabs.harvard.edu/abs/2017A&A...606L..12G}
}

@ARTICLE{2016JMoSp.327...95E,
       author = {{Endres}, Christian P. and {Schlemmer}, Stephan and {Schilke}, Peter and {Stutzki}, J{\"u}rgen and {M{\"u}ller}, Holger S.~P.},
        title = "{The Cologne Database for Molecular Spectroscopy, CDMS, in the Virtual Atomic and Molecular Data Centre, VAMDC}",
      journal = {Journal of Molecular Spectroscopy},
         year = 2016,
        month = sep,
       volume = {327},
        pages = {95-104},
          doi = {10.1016/j.jms.2016.03.005},
archivePrefix = {arXiv},
       eprint = {1603.03264},
 primaryClass = {astro-ph.IM},
       adsurl = {https://ui.adsabs.harvard.edu/abs/2016JMoSp.327...95E}
}

@ARTICLE{2022PASP..134k4501C,
       author = {{CASA Team} and {Bean}, Ben and {Bhatnagar}, Sanjay and {Castro}, Sandra and {Donovan Meyer}, Jennifer and {Emonts}, Bjorn and {Garcia}, Enrique and {Garwood}, Robert and {Golap}, Kumar and {Gonzalez Villalba}, Justo and {Harris}, Pamela and {Hayashi}, Yohei and {Hoskins}, Josh and {Hsieh}, Mingyu and {Jagannathan}, Preshanth and {Kawasaki}, Wataru and {Keimpema}, Aard and {Kettenis}, Mark and {Lopez}, Jorge and {Marvil}, Joshua and {Masters}, Joseph and {McNichols}, Andrew and {Mehringer}, David and {Miel}, Renaud and {Moellenbrock}, George and {Montesino}, Federico and {Nakazato}, Takeshi and {Ott}, Juergen and {Petry}, Dirk and {Pokorny}, Martin and {Raba}, Ryan and {Rau}, Urvashi and {Schiebel}, Darrell and {Schweighart}, Neal and {Sekhar}, Srikrishna and {Shimada}, Kazuhiko and {Small}, Des and {Steeb}, Jan-Willem and {Sugimoto}, Kanako and {Suoranta}, Ville and {Tsutsumi}, Takahiro and {van Bemmel}, Ilse M. and {Verkouter}, Marjolein and {Wells}, Akeem and {Xiong}, Wei and {Szomoru}, Arpad and {Griffith}, Morgan and {Glendenning}, Brian and {Kern}, Jeff},
        title = "{CASA, the Common Astronomy Software Applications for Radio Astronomy}",
      journal = {\pasp},
         year = 2022,
        month = nov,
       volume = {134},
       number = {1041},
          eid = {114501},
        pages = {114501},
          doi = {10.1088/1538-3873/ac9642},
archivePrefix = {arXiv},
       eprint = {2210.02276},
 primaryClass = {astro-ph.IM},
       adsurl = {https://ui.adsabs.harvard.edu/abs/2022PASP..134k4501C}
}

@ARTICLE{1994AA...291..943O,
       author = {{Ossenkopf}, V. and {Henning}, Th.},
        title = "{Dust opacities for protostellar cores.}",
      journal = {\aap},
         year = 1994,
        month = nov,
       volume = {291},
        pages = {943-959},
       adsurl = {https://ui.adsabs.harvard.edu/abs/1994A&A...291..943O}
}

@ARTICLE{2018ApJ...866..150T,
       author = {{Taniguchi}, Kotomi and {Saito}, Masao and {Majumdar}, Liton and {Shimoikura}, Tomomi and {Dobashi}, Kazuhito and {Ozeki}, Hiroyuki and {Nakamura}, Fumitaka and {Hirota}, Tomoya and {Minamidani}, Tetsuhiro and {Miyamoto}, Yusuke and {Kaneko}, Hiroyuki},
        title = "{Chemical Diversity in Three Massive Young Stellar Objects Associated with 6.7 GHz CH$_{3}$OH Masers}",
      journal = {\apj},
         year = 2018,
        month = oct,
       volume = {866},
       number = {2},
          eid = {150},
        pages = {150},
          doi = {10.3847/1538-4357/aade97},
archivePrefix = {arXiv},
       eprint = {1804.05205},
 primaryClass = {astro-ph.SR},
       adsurl = {https://ui.adsabs.harvard.edu/abs/2018ApJ...866..150T}
}

@ARTICLE{2009ARA&A..47..427H,
       author = {{Herbst}, Eric and {van Dishoeck}, Ewine F.},
        title = "{Complex Organic Interstellar Molecules}",
      journal = {\araa},
         year = 2009,
        month = sep,
       volume = {47},
       number = {1},
        pages = {427-480},
          doi = {10.1146/annurev-astro-082708-101654},
       adsurl = {https://ui.adsabs.harvard.edu/abs/2009ARA&A..47..427H}
}

@ARTICLE{1992ApJ...392..551S,
       author = {{Suzuki}, Hiroko and {Yamamoto}, Satoshi and {Ohishi}, Masatoshi and {Kaifu}, Norio and {Ishikawa}, Shin-Ichi and {Hirahara}, Yasuhiro and {Takano}, Shuro},
        title = "{A Survey of CCS, HC 3N, HC 5N, and NH 3 toward Dark Cloud Cores and Their Production Chemistry}",
      journal = {\apj},
         year = 1992,
        month = jun,
       volume = {392},
        pages = {551},
          doi = {10.1086/171456},
       adsurl = {https://ui.adsabs.harvard.edu/abs/1992ApJ...392..551S}
}

@ARTICLE{2008ApJ...672..371S,
       author = {{Sakai}, Nami and {Sakai}, Takeshi and {Hirota}, Tomoya and {Yamamoto}, Satoshi},
        title = "{Abundant Carbon-Chain Molecules toward the Low-Mass Protostar IRAS 04368+2557 in L1527}",
      journal = {\apj},
         year = 2008,
        month = jan,
       volume = {672},
       number = {1},
        pages = {371-381},
          doi = {10.1086/523635},
       adsurl = {https://ui.adsabs.harvard.edu/abs/2008ApJ...672..371S}
}

@ARTICLE{2008ApJ...681.1385H,
       author = {{Hassel}, George E. and {Herbst}, Eric and {Garrod}, Robin T.},
        title = "{Modeling the Lukewarm Corino Phase: Is L1527 Unique?}",
      journal = {\apj},
         year = 2008,
        month = jul,
       volume = {681},
       number = {2},
        pages = {1385-1395},
          doi = {10.1086/588185},
archivePrefix = {arXiv},
       eprint = {0803.1805},
 primaryClass = {astro-ph},
       adsurl = {https://ui.adsabs.harvard.edu/abs/2008ApJ...681.1385H}
}

@ARTICLE{2018ApJ...854..133T,
       author = {{Taniguchi}, Kotomi and {Saito}, Masao and {Sridharan}, T.~K. and {Minamidani}, Tetsuhiro},
        title = "{Survey Observations to Study Chemical Evolution from High-mass Starless Cores to High-mass Protostellar Objects. I. HC$_{3}$N and HC$_{5}$N}",
      journal = {\apj},
         year = 2018,
        month = feb,
       volume = {854},
       number = {2},
          eid = {133},
        pages = {133},
          doi = {10.3847/1538-4357/aaa66f},
archivePrefix = {arXiv},
       eprint = {1801.02116},
 primaryClass = {astro-ph.GA},
       adsurl = {https://ui.adsabs.harvard.edu/abs/2018ApJ...854..133T}
}

@ARTICLE{2019ApJ...872..154T,
       author = {{Taniguchi}, Kotomi and {Saito}, Masao and {Sridharan}, T.~K. and {Minamidani}, Tetsuhiro},
        title = "{Survey Observations to Study Chemical Evolution from High-mass Starless Cores to High-mass Protostellar Objects. II. HC$_{3}$N and N$_{2}$H$^{+}$}",
      journal = {\apj},
         year = 2019,
        month = feb,
       volume = {872},
       number = {2},
          eid = {154},
        pages = {154},
          doi = {10.3847/1538-4357/ab001e},
archivePrefix = {arXiv},
       eprint = {1901.06446},
 primaryClass = {astro-ph.GA},
       adsurl = {https://ui.adsabs.harvard.edu/abs/2019ApJ...872..154T}
}

@ARTICLE{2014MNRAS.443.2252G,
       author = {{Green}, C.-E. and {Green}, J.~A. and {Burton}, M.~G. and {Horiuchi}, S. and {Tothill}, N.~F.~H. and {Walsh}, A.~J. and {Purcell}, C.~R. and {Lovell}, J.~E.~J. and {Millar}, T.~J.},
        title = "{New detections of HC$_{5}$N towards hot cores associated with 6.7 GHz methanol masers}",
      journal = {\mnras},
         year = 2014,
        month = sep,
       volume = {443},
       number = {3},
        pages = {2252-2263},
          doi = {10.1093/mnras/stu1349},
archivePrefix = {arXiv},
       eprint = {1407.1699},
 primaryClass = {astro-ph.GA},
       adsurl = {https://ui.adsabs.harvard.edu/abs/2014MNRAS.443.2252G}
}

@ARTICLE{2017ApJ...844...68T,
       author = {{Taniguchi}, Kotomi and {Saito}, Masao and {Hirota}, Tomoya and {Ozeki}, Hiroyuki and {Miyamoto}, Yusuke and {Kaneko}, Hiroyuki and {Minamidani}, Tetsuhiro and {Shimoikura}, Tomomi and {Nakamura}, Fumitaka and {Dobashi}, Kazuhito},
        title = "{Observations of Cyanopolyynes toward Four High-mass Star-forming Regions Containing Hot Cores}",
      journal = {\apj},
         year = 2017,
        month = jul,
       volume = {844},
       number = {1},
          eid = {68},
        pages = {68},
          doi = {10.3847/1538-4357/aa7899},
archivePrefix = {arXiv},
       eprint = {1706.02465},
 primaryClass = {astro-ph.GA},
       adsurl = {https://ui.adsabs.harvard.edu/abs/2017ApJ...844...68T}
}

@ARTICLE{2008AJ....136.2391C,
       author = {{Cyganowski}, C.~J. and {Whitney}, B.~A. and {Holden}, E. and {Braden}, E. and {Brogan}, C.~L. and {Churchwell}, E. and {Indebetouw}, R. and {Watson}, D.~F. and {Babler}, B.~L. and {Benjamin}, R. and {Gomez}, M. and {Meade}, M.~R. and {Povich}, M.~S. and {Robitaille}, T.~P. and {Watson}, C.},
        title = "{A Catalog of Extended Green Objects in the GLIMPSE Survey: A New Sample of Massive Young Stellar Object Outflow Candidates}",
      journal = {\aj},
         year = 2008,
        month = dec,
       volume = {136},
       number = {6},
        pages = {2391-2412},
          doi = {10.1088/0004-6256/136/6/2391},
archivePrefix = {arXiv},
       eprint = {0810.0530},
 primaryClass = {astro-ph},
       adsurl = {https://ui.adsabs.harvard.edu/abs/2008AJ....136.2391C}
}

@ARTICLE{2009ApJ...702.1615C,
       author = {{Cyganowski}, C.~J. and {Brogan}, C.~L. and {Hunter}, T.~R. and {Churchwell}, E.},
        title = "{A Class I and Class II CH$_{3}$OH Maser Survey of EGOs from the GLIMPSE Survey}",
      journal = {\apj},
         year = 2009,
        month = sep,
       volume = {702},
       number = {2},
        pages = {1615-1647},
          doi = {10.1088/0004-637X/702/2/1615},
archivePrefix = {arXiv},
       eprint = {0907.1089},
 primaryClass = {astro-ph.SR},
       adsurl = {https://ui.adsabs.harvard.edu/abs/2009ApJ...702.1615C}
}

@ARTICLE{2003PASP..115..953B,
       author = {{Benjamin}, Robert A. and {Churchwell}, E. and {Babler}, Brian L. and {Bania}, T.~M. and {Clemens}, Dan P. and {Cohen}, Martin and {Dickey}, John M. and {Indebetouw}, R{\'e}my and {Jackson}, James M. and {Kobulnicky}, Henry A. and {Lazarian}, Alex and {Marston}, A.~P. and {Mathis}, John S. and {Meade}, Marilyn R. and {Seager}, Sara and {Stolovy}, S.~R. and {Watson}, C. and {Whitney}, Barbara A. and {Wolff}, Michael J. and {Wolfire}, Mark G.},
        title = "{GLIMPSE. I. An SIRTF Legacy Project to Map the Inner Galaxy}",
      journal = {\pasp},
         year = 2003,
        month = aug,
       volume = {115},
       number = {810},
        pages = {953-964},
          doi = {10.1086/376696},
archivePrefix = {arXiv},
       eprint = {astro-ph/0306274},
 primaryClass = {astro-ph},
       adsurl = {https://ui.adsabs.harvard.edu/abs/2003PASP..115..953B}
}

@ARTICLE{2023ApJ...942....7F,
       author = {{Fedriani}, Rub{\'e}n and {Tan}, Jonathan C. and {Telkamp}, Zoie and {Zhang}, Yichen and {Yang}, Yao-Lun and {Liu}, Mengyao and {De Buizer}, James M. and {Law}, Chi-Yan and {Beltran}, Maria T. and {Rosero}, Viviana and {Tanaka}, Kei E.~I. and {Cosentino}, Giuliana and {Gorai}, Prasanta and {Farias}, Juan and {Staff}, Jan E. and {Whitney}, Barbara},
        title = "{The SOFIA Massive (SOMA) Star Formation Survey. IV. Isolated Protostars}",
      journal = {\apj},
         year = 2023,
        month = jan,
       volume = {942},
       number = {1},
          eid = {7},
        pages = {7},
          doi = {10.3847/1538-4357/aca4cf},
archivePrefix = {arXiv},
       eprint = {2205.11422},
 primaryClass = {astro-ph.GA},
       adsurl = {https://ui.adsabs.harvard.edu/abs/2023ApJ...942....7F}
}

@ARTICLE{2025ApJ...986...15T,
       author = {{Telkamp}, Zoie and {Fedriani}, Rub{\'e}n and {Tan}, Jonathan C. and {Law}, Chi-Yan and {Zhang}, Yichen and {Plunkett}, Adele and {Crowe}, Samuel and {Yang}, Yao-Lun and {De Buizer}, James M. and {Beltran}, Maria T. and {Bonfand}, M{\'e}lisse and {Boyden}, Ryan and {Cosentino}, Giuliana and {Gorai}, Prasanta and {Liu}, Mengyao and {Rosero}, Viviana and {Taniguchi}, Kotomi and {Tanaka}, Kei E.~I. and {Rodr{\'\i}guez}, Tatiana M.},
        title = "{The SOFIA Massive (SOMA) Star Formation Survey. V. Clustered Protostars}",
      journal = {\apj},
         year = 2025,
        month = jun,
       volume = {986},
       number = {1},
          eid = {15},
        pages = {15},
          doi = {10.3847/1538-4357/adcd79},
archivePrefix = {arXiv},
       eprint = {2412.11792},
 primaryClass = {astro-ph.GA},
       adsurl = {https://ui.adsabs.harvard.edu/abs/2025ApJ...986...15T}
}

@ARTICLE{2004ApJS..154....1W,
       author = {{Werner}, M.~W. and {Roellig}, T.~L. and {Low}, F.~J. and {Rieke}, G.~H. and {Rieke}, M. and {Hoffmann}, W.~F. and {Young}, E. and {Houck}, J.~R. and {Brandl}, B. and {Fazio}, G.~G. and {Hora}, J.~L. and {Gehrz}, R.~D. and {Helou}, G. and {Soifer}, B.~T. and {Stauffer}, J. and {Keene}, J. and {Eisenhardt}, P. and {Gallagher}, D. and {Gautier}, T.~N. and {Irace}, W. and {Lawrence}, C.~R. and {Simmons}, L. and {Van Cleve}, J.~E. and {Jura}, M. and {Wright}, E.~L. and {Cruikshank}, D.~P.},
        title = "{The Spitzer Space Telescope Mission}",
      journal = {\apjs},
         year = 2004,
        month = sep,
       volume = {154},
       number = {1},
        pages = {1-9},
          doi = {10.1086/422992},
archivePrefix = {arXiv},
       eprint = {astro-ph/0406223},
 primaryClass = {astro-ph},
       adsurl = {https://ui.adsabs.harvard.edu/abs/2004ApJS..154....1W}
}

@ARTICLE{2004ApJS..154...10F,
       author = {{Fazio}, G.~G. and {Hora}, J.~L. and {Allen}, L.~E. and {Ashby}, M.~L.~N. and {Barmby}, P. and {Deutsch}, L.~K. and {Huang}, J.-S. and {Kleiner}, S. and {Marengo}, M. and {Megeath}, S.~T. and {Melnick}, G.~J. and {Pahre}, M.~A. and {Patten}, B.~M. and {Polizotti}, J. and {Smith}, H.~A. and {Taylor}, R.~S. and {Wang}, Z. and {Willner}, S.~P. and {Hoffmann}, W.~F. and {Pipher}, J.~L. and {Forrest}, W.~J. and {McMurty}, C.~W. and {McCreight}, C.~R. and {McKelvey}, M.~E. and {McMurray}, R.~E. and {Koch}, D.~G. and {Moseley}, S.~H. and {Arendt}, R.~G. and {Mentzell}, J.~E. and {Marx}, C.~T. and {Losch}, P. and {Mayman}, P. and {Eichhorn}, W. and {Krebs}, D. and {Jhabvala}, M. and {Gezari}, D.~Y. and {Fixsen}, D.~J. and {Flores}, J. and {Shakoorzadeh}, K. and {Jungo}, R. and {Hakun}, C. and {Workman}, L. and {Karpati}, G. and {Kichak}, R. and {Whitley}, R. and {Mann}, S. and {Tollestrup}, E.~V. and {Eisenhardt}, P. and {Stern}, D. and {Gorjian}, V. and {Bhattacharya}, B. and {Carey}, S. and {Nelson}, B.~O. and {Glaccum}, W.~J. and {Lacy}, M. and {Lowrance}, P.~J. and {Laine}, S. and {Reach}, W.~T. and {Stauffer}, J.~A. and {Surace}, J.~A. and {Wilson}, G. and {Wright}, E.~L. and {Hoffman}, A. and {Domingo}, G. and {Cohen}, M.},
        title = "{The Infrared Array Camera (IRAC) for the Spitzer Space Telescope}",
      journal = {\apjs},
         year = 2004,
        month = sep,
       volume = {154},
       number = {1},
        pages = {10-17},
          doi = {10.1086/422843},
archivePrefix = {arXiv},
       eprint = {astro-ph/0405616},
 primaryClass = {astro-ph},
       adsurl = {https://ui.adsabs.harvard.edu/abs/2004ApJS..154...10F}
}

@ARTICLE{2010A&A...518L...3G,
       author = {{Griffin}, M.~J. and {Abergel}, A. and {Abreu}, A. and {Ade}, P.~A.~R. and {Andr{\'e}}, P. and {Augueres}, J.-L. and {Babbedge}, T. and {Bae}, Y. and {Baillie}, T. and {Baluteau}, J.-P. and {Barlow}, M.~J. and {Bendo}, G. and {Benielli}, D. and {Bock}, J.~J. and {Bonhomme}, P. and {Brisbin}, D. and {Brockley-Blatt}, C. and {Caldwell}, M. and {Cara}, C. and {Castro-Rodriguez}, N. and {Cerulli}, R. and {Chanial}, P. and {Chen}, S. and {Clark}, E. and {Clements}, D.~L. and {Clerc}, L. and {Coker}, J. and {Communal}, D. and {Conversi}, L. and {Cox}, P. and {Crumb}, D. and {Cunningham}, C. and {Daly}, F. and {Davis}, G.~R. and {de Antoni}, P. and {Delderfield}, J. and {Devin}, N. and {di Giorgio}, A. and {Didschuns}, I. and {Dohlen}, K. and {Donati}, M. and {Dowell}, A. and {Dowell}, C.~D. and {Duband}, L. and {Dumaye}, L. and {Emery}, R.~J. and {Ferlet}, M. and {Ferrand}, D. and {Fontignie}, J. and {Fox}, M. and {Franceschini}, A. and {Frerking}, M. and {Fulton}, T. and {Garcia}, J. and {Gastaud}, R. and {Gear}, W.~K. and {Glenn}, J. and {Goizel}, A. and {Griffin}, D.~K. and {Grundy}, T. and {Guest}, S. and {Guillemet}, L. and {Hargrave}, P.~C. and {Harwit}, M. and {Hastings}, P. and {Hatziminaoglou}, E. and {Herman}, M. and {Hinde}, B. and {Hristov}, V. and {Huang}, M. and {Imhof}, P. and {Isaak}, K.~J. and {Israelsson}, U. and {Ivison}, R.~J. and {Jennings}, D. and {Kiernan}, B. and {King}, K.~J. and {Lange}, A.~E. and {Latter}, W. and {Laurent}, G. and {Laurent}, P. and {Leeks}, S.~J. and {Lellouch}, E. and {Levenson}, L. and {Li}, B. and {Li}, J. and {Lilienthal}, J. and {Lim}, T. and {Liu}, S.~J. and {Lu}, N. and {Madden}, S. and {Mainetti}, G. and {Marliani}, P. and {McKay}, D. and {Mercier}, K. and {Molinari}, S. and {Morris}, H. and {Moseley}, H. and {Mulder}, J. and {Mur}, M. and {Naylor}, D.~A. and {Nguyen}, H. and {O'Halloran}, B. and {Oliver}, S. and {Olofsson}, G. and {Olofsson}, H.-G. and {Orfei}, R. and {Page}, M.~J. and {Pain}, I. and {Panuzzo}, P. and {Papageorgiou}, A. and {Parks}, G. and {Parr-Burman}, P. and {Pearce}, A. and {Pearson}, C. and {P{\'e}rez-Fournon}, I. and {Pinsard}, F. and {Pisano}, G. and {Podosek}, J. and {Pohlen}, M. and {Polehampton}, E.~T. and {Pouliquen}, D. and {Rigopoulou}, D. and {Rizzo}, D. and {Roseboom}, I.~G. and {Roussel}, H. and {Rowan-Robinson}, M. and {Rownd}, B. and {Saraceno}, P. and {Sauvage}, M. and {Savage}, R. and {Savini}, G. and {Sawyer}, E. and {Scharmberg}, C. and {Schmitt}, D. and {Schneider}, N. and {Schulz}, B. and {Schwartz}, A. and {Shafer}, R. and {Shupe}, D.~L. and {Sibthorpe}, B. and {Sidher}, S. and {Smith}, A. and {Smith}, A.~J. and {Smith}, D. and {Spencer}, L. and {Stobie}, B. and {Sudiwala}, R. and {Sukhatme}, K. and {Surace}, C. and {Stevens}, J.~A. and {Swinyard}, B.~M. and {Trichas}, M. and {Tourette}, T. and {Triou}, H. and {Tseng}, S. and {Tucker}, C. and {Turner}, A. and {Vaccari}, M. and {Valtchanov}, I. and {Vigroux}, L. and {Virique}, E. and {Voellmer}, G. and {Walker}, H. and {Ward}, R. and {Waskett}, T. and {Weilert}, M. and {Wesson}, R. and {White}, G.~J. and {Whitehouse}, N. and {Wilson}, C.~D. and {Winter}, B. and {Woodcraft}, A.~L. and {Wright}, G.~S. and {Xu}, C.~K. and {Zavagno}, A. and {Zemcov}, M. and {Zhang}, L. and {Zonca}, E.},
        title = "{The Herschel-SPIRE instrument and its in-flight performance}",
      journal = {\aap},
         year = 2010,
        month = jul,
       volume = {518},
          eid = {L3},
        pages = {L3},
          doi = {10.1051/0004-6361/201014519},
archivePrefix = {arXiv},
       eprint = {1005.5123},
 primaryClass = {astro-ph.IM},
       adsurl = {https://ui.adsabs.harvard.edu/abs/2010A&A...518L...3G}
}

@ARTICLE{2018ApJ...853...18Z,
       author = {{Zhang}, Yichen and {Tan}, Jonathan C.},
        title = "{Radiation Transfer of Models of Massive Star Formation. IV. The Model Grid and Spectral Energy Distribution Fitting}",
      journal = {\apj},
         year = 2018,
        month = jan,
       volume = {853},
       number = {1},
          eid = {18},
        pages = {18},
          doi = {10.3847/1538-4357/aaa24a},
archivePrefix = {arXiv},
       eprint = {1708.08853},
 primaryClass = {astro-ph.GA},
       adsurl = {https://ui.adsabs.harvard.edu/abs/2018ApJ...853...18Z}
}

@ARTICLE{2022MNRAS.512.4419P,
       author = {{Peng}, Yaping and {Liu}, Tie and {Qin}, Sheng-Li and {Baug}, Tapas and {Liu}, Hong-Li and {Wang}, Ke and {Garay}, Guido and {Zhang}, Chao and {Chen}, Long-Fei and {Lee}, Chang Won and {Juvela}, Mika and {Li}, Dalei and {Tatematsu}, Ken'ichi and {Liu}, Xun-Chuan and {Lee}, Jeong-Eun and {Luo}, Gan and {Dewangan}, Lokesh and {Wu}, Yue-Fang and {Zhang}, Li and {Bronfman}, Leonardo and {Ge}, Jixing and {Tang}, Mengyao and {Zhang}, Yong and {Xu}, Feng-Wei and {Wang}, Yao and {Zhou}, Bing},
        title = "{ATOMS: ALMA Three-millimeter Observations of Massive Star-forming regions - X. Chemical differentiation among the massive cores in G9.62+0.19}",
      journal = {\mnras},
         year = 2022,
        month = may,
       volume = {512},
       number = {3},
        pages = {4419-4440},
          doi = {10.1093/mnras/stac624},
archivePrefix = {arXiv},
       eprint = {2203.12255},
 primaryClass = {astro-ph.GA},
       adsurl = {https://ui.adsabs.harvard.edu/abs/2022MNRAS.512.4419P}
}

@ARTICLE{2024MNRAS.533.1583L,
       author = {{Li}, Chuanshou and {Qin}, Sheng-Li and {Liu}, Tie and {Liu}, Sheng-Yuan and {Tang}, Mengyao and {Liu}, Hong-Li and {Chen}, Li and {Li}, Xiaohu and {Xu}, Fengwei and {Zhang}, Tianwei and {Liu}, Meizhu and {Shi}, Hongqiong and {Wu}, Yuefang},
        title = "{Correlations of methyl formate (CH$_{3}$OCHO), dimethyl ether (CH$_{3}$OCH$_{3}$), and ketene (H$_{2}$CCO) in high-mass star-forming regions}",
      journal = {\mnras},
         year = 2024,
        month = sep,
       volume = {533},
       number = {2},
        pages = {1583-1617},
          doi = {10.1093/mnras/stae1934},
archivePrefix = {arXiv},
       eprint = {2409.13192},
 primaryClass = {astro-ph.GA},
       adsurl = {https://ui.adsabs.harvard.edu/abs/2024MNRAS.533.1583L}
}

@ARTICLE{2011ApJ...739...63J,
       author = {{Juvela}, M. and {Ysard}, N.},
        title = "{On the Gas Temperature of Molecular Cloud Cores}",
      journal = {\apj},
         year = 2011,
        month = oct,
       volume = {739},
       number = {2},
          eid = {63},
        pages = {63},
          doi = {10.1088/0004-637X/739/2/63},
archivePrefix = {arXiv},
       eprint = {1108.1345},
 primaryClass = {astro-ph.GA},
       adsurl = {https://ui.adsabs.harvard.edu/abs/2011ApJ...739...63J}
}

@ARTICLE{2009MNRAS.394..221C,
       author = {{Chapman}, J.~F. and {Millar}, T.~J. and {Wardle}, M. and {Burton}, M.~G. and {Walsh}, A.~J.},
        title = "{Cyanopolyynes in hot cores: modelling G305.2+0.2}",
      journal = {\mnras},
         year = 2009,
        month = mar,
       volume = {394},
       number = {1},
        pages = {221-230},
          doi = {10.1111/j.1365-2966.2008.14144.x},
       adsurl = {https://ui.adsabs.harvard.edu/abs/2009MNRAS.394..221C}
}

@ARTICLE{2016ApJ...817..147T,
       author = {{Taniguchi}, Kotomi and {Ozeki}, Hiroyuki and {Saito}, Masao and {Sakai}, Nami and {Nakamura}, Fumitaka and {Kameno}, Seiji and {Takano}, Shuro and {Yamamoto}, Satoshi},
        title = "{Implication of Formation Mechanisms of HC$_{5}$N in TMC-1 as Studied by $^{13}$C Isotopic Fractionation}",
      journal = {\apj},
         year = 2016,
        month = feb,
       volume = {817},
       number = {2},
          eid = {147},
        pages = {147},
          doi = {10.3847/0004-637X/817/2/147},
archivePrefix = {arXiv},
       eprint = {1512.05783},
 primaryClass = {astro-ph.SR},
       adsurl = {https://ui.adsabs.harvard.edu/abs/2016ApJ...817..147T}
}

@ARTICLE{2018MNRAS.474.5068B,
       author = {{Burkhardt}, A.~M. and {Herbst}, E. and {Kalenskii}, S.~V. and {McCarthy}, M.~C. and {Remijan}, A.~J. and {McGuire}, B.~A.},
        title = "{Detection of HC$_{5}$N and HC$_{7}$N Isotopologues in TMC-1 with the Green Bank Telescope}",
      journal = {\mnras},
         year = 2018,
        month = mar,
       volume = {474},
       number = {4},
        pages = {5068-5075},
          doi = {10.1093/mnras/stx2972},
archivePrefix = {arXiv},
       eprint = {1711.07495},
 primaryClass = {astro-ph.GA},
       adsurl = {https://ui.adsabs.harvard.edu/abs/2018MNRAS.474.5068B}
}

@ARTICLE{2016ApJ...830..106T,
       author = {{Taniguchi}, Kotomi and {Saito}, Masao and {Ozeki}, Hiroyuki},
        title = "{$^{13}$C Isotopic Fractionation of HC$_{3}$N in Star-forming Regions: Low-mass Star-forming Region L1527 and High-mass Star-forming Region G28.28-0.36}",
      journal = {\apj},
         year = 2016,
        month = oct,
       volume = {830},
       number = {2},
          eid = {106},
        pages = {106},
          doi = {10.3847/0004-637X/830/2/106},
archivePrefix = {arXiv},
       eprint = {1608.01702},
 primaryClass = {astro-ph.GA},
       adsurl = {https://ui.adsabs.harvard.edu/abs/2016ApJ...830..106T}
}
\bibliographystyle{aasjournalv7}



\end{document}